\documentclass[12pt,preprint]{aastex}
\usepackage{url}
\usepackage[usenames]{color}
\usepackage{amsmath}
\usepackage{bm}
\usepackage[hyperindex,breaklinks]{hyperref}
\newcommand{\logg} {\log \textsl{\textrm{g}}}

\newcommand{\Te} {T_{\rm eff}}

\newcommand\gta{\lower 0.5ex\hbox{$\buildrel > \over \sim\ $}} 
\newcommand\lta{\lower 0.5ex\hbox{$\buildrel < \over \sim\ $}} 

\newcommand{\bra} {\langle}
\newcommand{\ket} {\rangle}

\begin{document}

\title{New Calculations of Stark Broadened Profiles for Neutral Helium Lines Using Computer Simulations}

\author{Patrick Tremblay, A. Beauchamp, and P. Bergeron}
\affil{D\'epartement de Physique, Universit\'e de Montr\'eal,
  C.P.~6128, Succ.~Centre-Ville, Montr\'eal, Qu\'ebec H3C 3J7, Canada}
\email{patrick@astro.umontreal.ca, bergeron@astro.umontreal.ca}

\begin{abstract}
We present new calculations of Stark broadened profiles for neutral
helium lines using computer simulations that include some important
aspects aimed at better representing the dynamical environment of the
helium atom. These include the unification of ion and electron
treatment, the correction for ion dynamics, the transition of the
electron contribution to broadening from the core to the wings of the
profile, the numerical integration of the time evolution operator of
helium perturbed by a fluctuating electric field, the Debye correction
for the correlation of the motion of charged perturbers, local density
variations, and particle reinjection. We compare the results of our
simulations for the He {\sc i} $\lambda\lambda$4471 and 4922 lines
with other results published in the literature. We also test our
simulation environment for narrow lines (He {\sc i}
$\lambda\lambda$5877 and 6678) and broader lines (He {\sc i}
$\lambda\lambda$4026 and 4144). We find that the narrow lines are more
difficult to produce adequately than the the broader ones.
\end{abstract}

\section{Introduction} \label{intro}

For nearly three decades, the atmospheric parameters of white dwarf
stars have been measured using the so-called spectroscopic technique,
where absorption line profiles are fitted with the predictions of
detailed model atmospheres (see, e.g,
\citealt{Bergeron92,Eisenstein2006,Kepler2007,Tremblay2011SDSS,Koester2015,GenestSDSS2019}). In
this case the effective temperature ($\Te$), surface gravity
($\logg$), and the chemical abundances can be measured, while the
other stellar parameters such as the stellar mass and radius, cooling
age, etc., can be derived from evolutionary models. Thanks to the
Sloan Digital Sky Survey (SDSS), atmospheric and physical parameters
have now been determined for tens of thousand white dwarfs \citep[see,
  e.g.,][]{Kepler2019}. While the spectroscopic technique is arguably
the most accurate approach to measure white dwarf parameters because
of the physical information contained in the line profiles, it is also
strongly dependent on the validity of the line profile calculations
(see \citealt{Tremblay2009} for the case of Stark broadening of
hydrogen lines). Also, the spectroscopic technique can only be applied
to white dwarfs that are hot enough to show strong absorption lines,
which corresponds roughly to $\Te>6500$~K for DA stars, and
$\Te>13,000$~K for DB stars. Thus white dwarfs close to the peak of
the white dwarf luminosity function cannot be studied using this
method.

An alternative approach that can be applied to any type of white dwarf
with relative accuracy is the photometric technique \citep{BRL97},
where the overall spectral energy distribution built from observed
magnitudes (converted into average fluxes) are fitted with synthetic
photometry obtained from model spectra. In this case, the effective
temperature and the solid angle --- $\pi(R/D)^2$ where $R$ is the
stellar radius and $D$ the distance from Earth --- are considered free
parameters, which implies that in order to measure the radius (and
other stellar parameters such as the mass), the distance to each
object must be known from trigonometric parallax measurements. The
photometric technique has been applied by \citet{BLR01} to nearly all
white dwarfs with trigonometric parallaxes available at that time,
which unfortunately, included only a few hundred objects. Also, the
optical $BVRI$ and infrared $JHK$ photometry they used was
painstakingly secured object by object. Now thanks to the recent
astrometric {\it Gaia} mission \citep{GaiaHRD}, precise astrometric
and photometric data for $\sim$260,000 high-confidence white dwarf
candidates have now become available. Furthermore, large (almost)
all-sky surveys are now providing exquisite photometric data, for
instance $ugriz$ photometry from SDSS, or $grizy$ photometry from
Pan-STARRS (Panoramic Survey Telescope And Rapid Response System;
\citealt{Tonry2012}).

Recently, \citet[][see also \citealt{GenestDB2019} for a more thorough
analysis of DB stars]{GenestSDSS2019} presented a detailed
spectroscopic and photometric analysis of DA and DB white dwarfs drawn
from the SDSS with trigonometric parallax measurements available from
the {\it Gaia} mission. While in principle both sets of parameters
obtained from spectroscopy and photometry should agree, large
discrepancies have been identified in terms of effective temperature
and mass determinations (or equivalently, $\logg$). Although
uncertainties in the treatment of line broadening theory, convective
energy transport, and perhaps chemical composition have been invoked
to account for these discrepancies, it is also possible that
calibration problems with the photometric and/or spectroscopic data
play an important role \citep{Bergeron2019}. In order to understand
better the nature of these discrepancies, every specific ingredient of
both the spectroscopic and photometric techniques needs to be
scrutinized in detail, both theoretically and observationally. In this
paper, we focus our attention on the Stark broadening of neutral
helium lines used in model atmosphere calculations of DB/DO white
dwarfs.

The main line broadening mechanism in hot ($\Te\gtrsim 15,000$~K) DB
white dwarfs is Stark broadening, which is caused by the presence of
charged perturbers in the surrounding plasma. Historically, the
neutral helium lines were treated within the semi-classical theory of
overlapping lines introduced by \citet{Baranger58c,Baranger58a}, which
consists of considering classical perturbers around a helium atom
described by quantum mechanics. A few detailed line profile
calculations became available where the quasi-static approximation was
used for the ions, together with the impact approximation for the
electrons; in these calculations, the influence of neighboring energy
levels to the observed transition was considered, giving rise to the
so-called forbidden components. These calculations include He {\sc i}
$\lambda$4471 \citep{Gieske69,BCS69}, He {\sc i} $\lambda$4922
\citep{BCS69}, He {\sc i} $\lambda$5016 \citep{BC70}, He {\sc i}
$\lambda$4388 \citep{Shamey69}, He {\sc i} $\lambda$4026
\citep{Shamey69,Gieske69}, and He {\sc i} $\lambda$6678
\citep{Ya72}. \cite{BCS74} and \cite{BCS75} then computed a new series
of line profiles for $\lambda$4471 and $\lambda$4922, respectively, by
including the one-electron approximation from \citet{Baranger62},
which provides a better transition between the impact approximation
and the quasi-static contribution of the electrons, along with ion
dynamics corrections. Unfortunately, the other helium lines in the
optical were still calculated using the isolated line approximation of
the theory, thus neglecting the contributions of the forbidden
components to the line profile (used by \citealt{Griem74} and
\citealt{Dimitri84}). The limited access to detailed neutral helium
line profiles in the optical proved to be a major obstacle to perform
reliable spectroscopic analyses of DB white dwarfs.

To remedy this situation, \citet[][see also
  \citealt{AlainPhD}]{beauchamp97} produced detailed line profile
calculations for most neutral helium lines in the optical using the
same semi-classical approximation as described above, by also
including the one-electron approximation while the ions were still
considered within the quasi-static approximation. These detailed line
profiles significantly improved the spectroscopic analyses of DB white
dwarfs (\citealt{Beauchamp1996}, \citealt{Voss2007},
\citealt{Bergeron2011}, \citealt{Koester2015},
\citealt{GenestDB2019}), and they remain the ultimate reference in the
field as of today. However, the downside of the semi-analytical
approach used by Beauchamp et al.~is the theoretical limitations
describing the analytical contribution to the line profile from both
the ions and the electrons. Even if there are several approximations
for the electrons --- namely the impact and one-electron
approximations --- for different regimes of interest, the smooth
transition between these two is not guaranteed. Also, the quasi-static
approximation becomes problematic in the line cores where ion dynamics
contribute, in particular at lower electron densities
\citep{BCS74,BCS75,Gig96,Omar06,SahalBrechot07,Gig14}.

An alternative approach to the semi-analytical calculations performed
by \citet{beauchamp97} is to produce full-fledged computer
simulations. This approach has first been introduced by
\citet{Stamm79} to study the effect of ion dynamics on the Ly$\beta$
hydrogen line. The method consists in generating a time sequence of
electric fields to represent their statistical distributions in a
natural way using a simple plasma model formed by charged particles
(perturbers) and emitters (neutral or charged). Although the method
appeared like an “ideal experiment”, computational time was the
largest obstacle, as well as the crude treatment of the limited
simulated volume. It is not until \citet{Gig85} that the computing
time was reduced using an improved differential equation solving
algorithm for the evolution of the emitter for hydrogen. This allowed
the calculations of spectra of hydrogen Balmer lines, and Stark
broadening could now be considered in more detail using a new particle
re-injection technique \citep{Gig87, Gig96}, in combination with a
more adequate spherical simulation volume, instead of a cubic
volume. It was also demonstrated that one could remove the limitations
of a finite simulation volume \citep{H88}. Improvements to these
techniques in the following years
\citep{Calisti88,Frerich89,Poquerusse96,Halenka96,Alex99v2,Alex99,Sorge00,Olcha02,Wu02,Halenka02,Wu03,Olcha04,Stambulchik2006,Stambulchik2007,Gig18}
allowed the calculations of both hydrogen \citep{Gig96,Gig03,GomezPhD}
and neutral helium \citep{Gig09,Lara12} Stark broadened line
profiles. The simulation method is now gaining more and more attention
\citep{Gig14}, even though the computational time remains the largest
obstacle as we gradually converge towards a more physically
representative simulation \citep{Gig18}.

We report in this paper the results of our own computer simulations,
performed by combining the method developed by \citet{Gig96} for
treating naturally the dynamics of the environment around the emitting
neutral helium atom, and the approach of \citet{H88} for dealing with
local density variations. The theoretical background and the standard
Stark broadening theory are first described in Sections \ref{sec:THEO}
and \ref{sec:TST}, respectively. We then discuss in Section
\ref{sec:simu} the details of our computer simulations, including the
generation and reinjection of particles in the simulation volume, and
the time evolution operator. We finally present in Section
\ref{sec:comp} our results for the He {\sc i} $\lambda\lambda$4026,
4144, 4471, 4922, 5877, and 6678 lines, and compare our line profiles
with those published by \citet{beauchamp97}, \citet{Gig09}, and
\citet{Lara12}; comparisons of synthetic spectra of DB white dwarfs
for He {\sc i} $\lambda$4471 and $\lambda$4922 are also discussed. Our
conclusions follow in Section \ref{sec:conc}.

\section{Theoretical Background}\label{sec:THEO}

\subsection{Fundamental Aspects of the Stark Broadening Theory}\label{sec:fundamentals}

Most Stark-broadened line profile calculations for He {\sc i} lines
are based on the semi-classical theory of Stark broadening of
\citet[][see also \citealt{Baranger62,Sahal69a,
    Smith69,Griem74,Gig14}]{Baranger58a}. The charged particles,
electrons and ions, which perturb the emitting helium atom move along
classical paths that are independent of the state of the atom, while
the spectral distribution of the light emitted by the atom is computed
quantum-mechanically.  Since the model is developed within the
coordinate system of the emitter, the effects induced by the motion of
the emitters relative to the observer are not taken into account, and
the computed Stark profile must thus subsequently be convoluted with a
Gaussian Doppler profile.

The development of the Stark broadening theory starts with an isolated
quantum system in thermodynamic equilibrium, composed of a single
emitter at rest interacting with charged perturbers with no internal
structure. The type and number density of the perturbers are
consistent with the given macroscopic properties of the gas
(temperature, electron number density, and chemical composition). The
Hamiltonian of the system can be written as

\begin{equation}
\label{eq:H}
    H=H_0+H_p+V\,
\end{equation}

\noindent
where $H_0$ is the Hamiltonian of the isolated emitter, $H_p$ is the
Hamiltonian of the perturbers, including their mutual interactions,
and $V$ is the interaction between the neutral emitter and the
perturbers.  Since the system is in thermodynamic equilibrium, the
Hamiltonian has no explicit time dependence, which implies that its
eigenstates have well-defined energies.

The power spectrum $P(\omega)$ of the plasma is defined as the power
emitted per unit frequency interval. The expression for the
spontaneous emission rate of an isolated atom can be generalized for
the whole quantum system as a weighted sum over all possible
spontaneous transitions $|i\ket\rightarrow|f\ket$ between initial and
final eigenstates of the Hamiltonian (Equation \ref{eq:H}),

\begin{equation}
\label{eq:Pomega}
    P(\omega)  =\frac{4\pi \omega^4}{3c^3}\sum_{if} \delta(\omega-\omega_{if})|\langle f|\mathbf{d}|i\rangle|^2 \rho_i\ ,
\end{equation}

\noindent where $\omega_{if}=(E_i-E_f)/\hbar$ is the angular frequency
of the emitted photon, $\mathbf{d}$ is the electric dipole moment
operator of the whole system, and $\rho_i$ is the statistical weight
of the initial state of the transition, i.e., its Boltzmann factor
given by

\begin{equation}
\label{eq:rho_i}
\rho_i \propto e^{-E_i/kT}\ ,
\end{equation}

\noindent whose sum over all initial states is normalized to unit area.

Since we only calculate the spectrum of the emitter, the contribution
of the perturbers to the dipole moment is generally omitted. In the
particular case where the emitter is a helium atom, all the lines in
the optical region of the spectrum are the result of transitions
between singly excited states, that is, with one electron staying in
the ground state. The dipole moment then reduces to that of the single
electron in the excited state,

\begin{equation}
{\bf d} = e{\bf R}\ ,    
\end{equation}

\noindent where $e$ is the elementary charge, and ${\bf R}$ the position operator of the electron.
 
One generally discards the physical constants and the slowly varying
$\omega^4$ term in Equation (\ref{eq:Pomega}), and redefines the
profile as

\begin{equation}
\label{eq:Fomega}
    I(\omega)  =\sum_{if} \delta(\omega-\omega_{if})|\langle f|\mathbf{d}|i\rangle|^2 \rho_i\ .
\end{equation}

\noindent In practice, it is usually easier to work in the time
domain, and to compute directly the Fourier transform of the profile,
called the autocorrelation function

\begin{equation}
\label{eq:Ct}
C(t)= \int_{-\infty}^{+\infty} dt\ e^{-i\omega t} I(\omega) =\sum_{if} e^{-i\omega_{if}t}|\langle f|\mathbf{d}|i\rangle|^2 \rho_i\ ,
\end{equation}

\noindent where $C(t)$ satisfies the condition 
$C(-t)=C^*(t)$ since $I(\omega)$ is real. The profile can then be recovered by performing the inverse Fourier transform

\begin{equation}
\label{eq:stark9}
    I(\omega)=\frac{1}{\pi}\mathrm{Re}\int_0^\infty dt\,e^{i\omega t} C(t)\ ,
\end{equation}

\noindent where the above condition on $C(t)$ is explicitly taken into
account. The theoretical development now focuses on the detailed
calculations of $C(t)$ from the dynamics of the system.

Equation (\ref{eq:Ct}) is tied to a particular basis, the energy
eigenstates of the whole system. It is thus advantageous to express
$C(t)$ with quantum operators only, resulting in a form independent of
the basis. This allows further theoretical development and numerical
computations to be performed in the most suited basis for the
task. Two terms must be converted into matrix elements: the
statistical weight and the oscillatory term. The former is expressed
as the diagonal element of the density matrix,

\begin{equation}
\rho_i = \bra i | \rho |i\ket\ ,
\end{equation}

\noindent where the operator

\begin{equation}
    \rho = e^{-H/kT}/\ {\rm Tr}\ (e^{-H/kT})
\end{equation}

\noindent is diagonal in the energy eigenstates basis. The second term
can be written in terms of the time evolution operator $U(t, 0)$,
which satisfies the Schr\"odinger equation

\begin{equation}
\label{eq:U}
   i\hbar\, \frac{dU}{dt} = HU\ 
\end{equation}
and the boundary condition $U(0, 0)=I$. Since $H$ does not depend
explicitly on time in Equation (\ref{eq:H}), the integration of
Equation (\ref{eq:U}) yields the trivial solution $U(t, 0)=
e^{-iHt/\hbar}$, which implies that
\begin{equation}
U|i\ket = e^{-iE_it/\hbar}|i\ket\ ,\ \ \ \ \ \ 
U^\dag|f\ket = e^{iE_ft/\hbar}|f\ket\ . 
\end{equation}
$C(t)$ can now be written as a trace over a product of quantum
operators, a form that is independent of the representation of the
states of the system
\begin{equation}
\begin{aligned}
\label{eq:autocorrel2article}
    C(t)  &=\sum_{if} e^{iE_f t/\hbar}\langle i|{\bf d}|f\rangle 
    e^{-iE_it/\hbar}\langle f|{\bf d}^\dagger|i\rangle \rho_i\\
    &=\sum_{if} 
    \langle i|{\bf d} U^\dag|f\rangle
    \langle f|{\bf d}^\dagger U|i\rangle \rho_i\\
    &=\mathrm{Tr}( \textbf{d}(U^{\dag}\,\textbf{d}^\dagger\,U\,)  \rho )\ .
\end{aligned}
\end{equation}

\subsection{Classical Path Approximation}\label{sec:classicalpath}

The classical path method divides the system into two subsystems, the
emitter and the perturbers, and treats them separately, making two
approximations that disentangle the two sets of coordinates in
Equation (\ref{eq:autocorrel2article}).

The weak coupling approximation assumes that the interaction potential
$V$ between the two subsystems is negligible relative to the kinetic
energy of a perturber \citep{Smith69}. The state of the perturbers
remains independent of the state of the emitter despite the
interaction that produces line broadening (``no backreaction"). This
has two consequences. First, the density matrix of the system becomes
separable

\begin{equation}
 \rho \propto  
 e^{-(H_0 + H_p + V)/kT} \approx e^{-H_0/kT}e^{-H_p/kT}
 \equiv \rho_a \rho_p\ ,
\end{equation}

\noindent where $\rho_a$ and $\rho_p$ only depend on the coordinates
of the atom and the perturbers, respectively. Next, the wave function
of the system can be approximated by the product of the perturbers and
emitter wave functions, the latter being subject to a time-dependent
interaction $V(t)$, with a Hamiltonian written as

\begin{equation}
\label{eq:ha}
H_a = H_0 + V(t)\ .    
\end{equation}

\noindent The autocorrelation function can now be written as two
embedded traces over the perturbers and atomic coordinates

\begin{equation}
\label{eq:C2}
    C(t)   =\mathrm{Tr_p}\{\  \mathrm{Tr_a}(\textbf{d}(U_a^{\dag}\,\textbf{d}^\dagger\,U_a\,)  \rho_a) \ \rho_p \}\,,
\end{equation}

\noindent where $U_a$ is the time evolution operator of the emitter, which satisfies the equation

\begin{equation}
\label{eq:Ua}
   i\hbar\,{dU_a\over dt}  = H_a U_a\ .
\end{equation}

The perturbers are still treated as quantum particles in Equation
(\ref{eq:C2}). The last step of the semi-classical approach is to
approximate the wave function of the perturbers by a product of wave
packets that are assumed to remain sufficiently localized during the
characteristic timescale of the broadening process. This allows the
replacement of $V(t)$ by the corresponding classical potential
function $V({\bf R}, {\bf x}(t))$, where ${\bf R}$ is the position
operator of the radiating electron and {\bf x} is the set of classical
3D-coordinates of all perturbers.

The perturbers now follow classical trajectories with well-defined
positions and velocities, and the trace over the states of the
perturbers in Equation (\ref{eq:C2}) is reinterpreted as an average
over all possible configurations of perturbers, called the thermal
average

\begin{equation}
\label{eq:C5}
    C(t) =\{\ \mathrm{Tr}_{a}( \textbf{d} \, U_a^\dagger\,\textbf{d}^\dagger\,U_a \rho_a )\ \}_{\mathrm{av}}\,,
\end{equation}

\noindent where the weight of a configuration of perturbers of energy
$E_p$ is a Boltzmann factor proportional to $e^{-E_p/kT}$. In most He {\sc i} 
line profile calculations, the interaction between perturbers is
neglected, and the statistical weight is computed from their kinetic
energies only, resulting in a product of Maxwell velocity
distributions. By separating the behavior of the classical perturbers
from the state of the atom, a series of dynamic configurations of
perturbers can then be defined, at least conceptually, prior to
quantum mechanical calculations.

\subsection{Calculation of the Autocorrelation Function}\label{sec:autocorrel}
 
Equation (\ref{eq:C5}) is still not applicable for practical
calculations since it includes the contribution of all possible
electronic transitions of the atom, and involves products of matrices
of infinite dimension. The trace being independent of the choice of
basis, we adopt the states of the isolated helium atom as the basis,
and express the autocorrelation function as a product of 5 matrices:

\begin{equation}
\label{eq:C6}
    C(t) =\ \{ \sum_{abb'a'a''} 
    \langle a|\mathbf{d}|b\rangle\ 
    \langle b|U^\dagger(t) |b'\rangle\ \langle b'|\mathbf{d}^\dagger|a'\rangle\ \langle a''|U(t, 0) |a''\rangle\ \langle a''| \rho|a\rangle
    \ \}_{\mathrm{av}}\ ,
\end{equation}

\noindent where every summation is formally performed over all
unperturbed states of the atom, and $U$ (with the $a$ subscript
omitted for clarity) now stands for the time evolution operator of the
atom only. The dummy indices in the summation are limited to finite
sets by the following arguments.
 
Every He {\sc i} line in the optical and in the near UV range is the
result of transitions between upper levels $|n\ell ms\ket$ with $n \ge
3$ and lower levels $|2\ell'm's\ket$ of the same spin $s$.  For the
calculation of a single line profile, the sets of $|a\ket$ and
$|b\ket$ states are defined, respectively, as the upper and lower
states of the transition under study.  The states within each group
share the same $n$, $\ell$, and $s$ quantum numbers, and differ only
in their $m$ values.

In this work, we neglect the perturbation of the lower levels, which
are in general much less polarizable than the upper levels. This
implies that $\bra b|U^\dag|b'\ket = e^{iE_bt/\hbar}\delta_{bb'}$, and
the $b'$ index can be removed from the summation in Equation
(\ref{eq:C6}). The density matrix $\rho$ is further assumed to be
diagonal in the unperturbed He {\sc i} states basis, an approximation
always used in practice in the calculations of He {\sc i} profiles, to
our knowledge.  Moreover, $\bra a|\rho|a\ket$ has the same value for
all $|a\ket$ states, which allows us to discard this multiplicative
factor because the final profile will be normalized to unity.
Combining these results, we find that

 \begin{equation}
\label{eq:C7}
    C(t) = \ \{ \ \sum_{aba'} e^{iE_bt/\hbar}
    \langle a|\mathbf{d}|b\rangle  
    \ \bra b|\mathbf{d}^\dagger|a'\rangle\ \langle a'|U(t, 0) |a\rangle\ \}_{\mathrm{av}}\ .
\end{equation}

\noindent As discussed in Appendix \ref{sec:per}, the upper states
$|a\ket$, when subjected to the potential $V(t)$, mix with their
neighboring states of the same spin, mainly those of the same quantum
number $n$. In this work, the $|a'\ket$ states will be defined as the
subspace of states with the same value of $n$ for the upper states of
the transition, a special case of the no-quenching approximation
\citep{Kolb58}, which assumes that collisions do not cause transitions
between states with different values of $n$. Thus, the $|a\ket$ states
are a subset of the $|a'\ket$ set of states, which is disjoint from
the lower $|b\ket$ states. For example, considering the $2P^1-4D^1$
$\lambda$4922 transition, the three sets of states are the seven
states $|4,\ell=2,m, s=0\ket$ with $|m|\le 2$, the 16 states $|4\ell
m, s=0\ket$ with $\ell \le 3$ and $|m|\le \ell$, and the single state
$|2,\ell=1,m,s=0\ket$ with $|m| \le 1$.

Our last approximation is the definition of the potential $V(t)$ in
the definition of the time evolution operator of the atom in Equation
(\ref{eq:Ua}). We only keep the first non-vanishing term in the
multipole expansion of the interaction between the atom and the
perturbers, which is the dipole term, which yields the Hamiltonian

\begin{equation}
\label{eqn:H_a}
    H_a = H_0 + e{\bf F}(t)\cdot {\bf R}\ ,
\end{equation}

\noindent where ${\bf F}$ is the electric field at the position of the emitter. 

At this stage, the standard theory and the computer simulation
approach take different routes for the integration of the Schr\"odinger
equation of the time evolution operator of the atom (Equation
\ref{eq:Ua}), and for the calculations of the thermal average
$\{...\}_{\rm av}$ over all configurations of classical perturbers.

\section{Standard Stark Broadening Theory}\label{sec:TST}

This section presents a brief outline of the standard Stark broadening
theory, and focuses on the core approximations of this theory, most of
which are corrected by the computer simulation method.
 
\subsection{Basic Principles \label{sec:principles}}

The standard Stark broadening theory uses the quasi-static
approximation for the ions and the impact approximation for the
electrons. This section presents a summary of this theory for the
simpler case where the lower levels of the transition are not
significantly perturbed (see \citealt{Baranger58a,Baranger62},
\citealt{Smith69B}, \citealt{Griem74}, \citealt{BCS69}, and
\citealt{Sahal69a,Sahal69b}, for a more general description of the
theory).

Since the ions are assumed to be quasi-static, effects stemming from
their motion are neglected, and the electric field associated with the
presence of ions may be chosen to point arbitrarily toward the $z$
direction. For each fixed configuration of ion perturbers, the
Hamiltonian of the emitter immersed in the field $F$ produced by the
ions then becomes

\begin{equation}
\label{eq:Ht}
H(t) = H_a + eFZ + V_e(t) 
\end{equation}

\noindent where $V_e$ is the time-varying potential due to the electrons, and $Z$ is the $z$-component of ${\bf R}$.

The interaction between ions and electrons is neglected, thus allowing
the separation of the thermal average into averages over
configurations of ions and electrons. Moreover, since the Hamiltonian
(Equation \ref{eq:Ht}) does not depend on the detailed configuration
of the ions, except through the ion electric field, the average over
the static ion configurations can be replaced by an average over ion
electric fields, weighted by the microfield statistical
distribution. The autocorrelation function of the profile then becomes

\begin{equation}
\label{eqn:CtStandard}
C(t) = \int_0^\infty dF \ W_R(F)\  \{C_e(t;F)\}_{{\rm av}_e}
\end{equation}

\noindent where $C_e(t;F)$ is the autocorrelation function due to the
moving electrons in the static field $F$. In this expression, the
distribution of the microfield $ W_R(F)$ takes into account the
correlation in the spatial distribution of the ions due to their
mutual interactions, and is parameterized by $R$, the ratio of the
mean distance between ions and the Debye radius \citep{Hoop68}.

The autocorrelation function $C_e(t;F)$ for the electron is a
generalization of Equation (\ref{eq:C7}), now applied to the
Hamiltonian in Equation (\ref{eq:Ht})

\begin{equation}
\label{eq:C10}
    C_e(t;F) = \sum_{\alpha b \alpha'} e^{iE_bt/\hbar}
    \langle \alpha|\mathbf{d}|b\rangle\ \bra b|\mathbf{d}^\dagger|\alpha'\rangle\ \{\ \langle \alpha'|U_e(t;F) |\alpha\rangle\ \}_{{\rm av}_e}\ ,
\end{equation}

\noindent where the summation is performed over all perturbed upper
states $|\alpha(F)\ket$ of the helium atom subjected to the constant
field $F$. Similarly, the time evolution operator satisfies the
Schr\"odinger equation

\begin{equation}
\label{eq:UaF}
   i\hbar\,{dU_e(t;F)\over dt}  = (H_a+eFZ+V_e(t))\, U_e(t;F)\ .
\end{equation}

\noindent Finally, the ``electronic" thermal average $\{...\}_{{\rm
    av}_e}$ is performed over all possible trajectories of the
electrons.

\subsection{The Impact Approximation \label{sec:impact}}

The electronic evolution operator $U_e(t;F)$ that satisfies the
Schr\"odinger equation (Equation \ref{eq:UaF}) can be formally expanded
as a Dyson series using perturbation theory,

\begin{equation}
\label{eq:UDyson}
U_e(t;F)=e^{-iH_a(F)t}\Bigg[1-{i\over \hbar}\int_0^t dt_1\,V'(t_1)+\Big(-{i\over \hbar}\Big)^2\int_0^t dt_1 \int_0^{t_1} dt_2\,V'(t_1)V'(t_2)+...\Bigg]
\end{equation}

\noindent where the operator $V'(t)\equiv
e^{-iH_a(F)t}V_e(t;F)e^{iH_a(F)t}$. The increasing complexity of the
successive terms in the expansion makes the handling of overlapping
collisions intractable, as well as the calculation of the contribution
of a single collision.

The impact approximation avoids these problems by assuming that (1)
strong collisions do not overlap in time, (2) the contribution of weak
overlapping collisions can be approximated by the truncated expansion
(Equation \ref{eq:UDyson}) up to second order, and (3) the collision
time is much shorter than the inverse half width at half maximum of
the line (the completed collision assumption). Performing the thermal
average of the resulting $U_e$ in Equation (\ref{eq:C10}), and
computing the inverse Fourier transform (Equation \ref{eq:stark9}),
then yield the line profile

\begin{equation}
\label{eq:QS}
    I(\omega)  = \int_0^\infty dF \ W_R(F) \ I_e(\omega;F)\ ,
\end{equation}

\noindent which is a weighted sum of electron-broadened profiles at constant ionic field

\begin{equation}
\label{eq:Ie}
I_e(\omega;F) = -{1\over \pi}\sum_{\alpha b \alpha'}
\bra \alpha |{\bf d}|b\ket
\bra b|{\bf d^\dag}|\alpha'\ket 
\bra \alpha' |\ [i(\omega-\omega_{\alpha b})+\Phi\ ]^{-1}|\alpha'\ket\ ,
\end{equation}

\noindent where $\hbar\omega_{\alpha b}$ is the energy difference
between the perturbed upper states $|\alpha\ket$ and unperturbed lower
states $|b\ket$, and the collision operator $\Phi$ represents the
contribution of the electrons (in the impact regime) to
broadening. Here, the dependence of the perturbed upper states, the
perturbed energies, and the collision operator $\Phi$ on the field $F$
is omitted. The diagonalization process that yields the perturbed
energies, as well as the unitary transformation between perturbed and
unperturbed states, are discussed in Appendix \ref{sec:per}.

As a result of the impact approximation, the collision operator $\Phi$
for a given electron number density $N_e$ corresponds to the second
order term of the expansion in Equation (\ref{eq:UDyson}) integrated
over the full trajectory of a single electron, and averaged over all
possible trajectories

\begin{equation}
\label{eq:UDyson2}
\Phi(F) = N_e \Big(-{i\over \hbar}\Big)^2\ \bigg\{\int_{-\infty}^{\infty} dt_1 \int_{-\infty}^{t_1} dt_2V'(t_1)V'(t_2) \bigg\}_{{\rm av}_e}.
\end{equation}

\noindent In this averaging process, the electron is assumed to move
along a rectilinear trajectory at constant velocity, described by the
expression

\begin{equation}
{\bf r}(t)={\bm{\rho}} + {\bf v}(t-t_0)\ ,
\end{equation}

\noindent where ${\bm{\rho}}$ and $t_0$ are the position and time of
nearest approach, respectively, and ${\bf v}$ is the velocity
perpendicular to ${\bm{\rho}}$. The impact parameter of the trajectory
is the norm of ${\bm{\rho}}$.

The averaging process involves integrals over the impact parameter
$\rho \equiv |{\bm{\rho}}|$, the orientation of the ${\bm{\rho}}$ and
${\bf v}$ vectors, as well as a weighted average over the
Maxwell-Boltzmann velocity distribution. The integral over $\rho$ is
formally between 0 and infinity, but upper and lower cut-offs are
defined for practical calculations. A maximal impact parameter
$\rho_{\rm max}$, of the order of the Debye radius, is used to
approximate the screening effect due to the correlation in the spatial
distribution of the electrons. A minimal impact parameter $\rho_{\rm
  min,\,\alpha\alpha'}(v;F)$ --- a function of the velocity, different
for each pair of states --- is further required because the
perturbation expansion (Equation \ref{eq:UDyson}) breaks down at small
impact parameters, which produces a diverging integral. The collision
operator $\Phi$ is therefore computed as a sum of contributions from
weak and strong collisions, characterized by impact parameters larger
and smaller than $\rho_{\rm min}$, respectively,

\begin{equation}
    \Phi(F) = \Phi(F)^{\rm weak} + \Phi(F)^{\rm strong}\ .
\end{equation}

\noindent The strong-collision term is calculated with a classical
theory \citep{Lorentz1906}, which is probably correct within a factor
2 \citep{Griem62}.

In the far wings, the complete collision approximation of the impact
theory is no longer valid. A unified theory that properly makes the
transition from the impact to the quasi-static regime for the electron
has been developed for hydrogen, taking advantage of the
$\ell$-degeneracy of the levels (\citealt{VCS70}). There is however no
such theory for the helium atom, and the impact broadening theory must
therefore be replaced by the one-electron theory.

\subsection{The One-Electron Theory \label{sec:oneelec}}

The one-electron theory of \citet{Baranger62} describes the profile in
the line wings, where non-overlapping strong collisions contribute,
and it also explains how the profile approximately turns into a
quasi-static line profile. This theory is not applicable to the entire
profile, however, and it even diverges in the line core. The
one-electron theory is not formally considered as part of the standard
Stark broadening theory, but it constitutes nevertheless the
alternative in the far wing where the impact approximation is no
longer valid. Fortunately, the interval of validity of the
one-perturber approximation partially overlaps with that of the impact
theory. The one-perturber theory describes the profile in the wings
where $\Delta \omega$ is much larger than the width of the line, while
the impact theory is valid when $\Delta \omega$ is less than a typical
collision time.

In the one-perturber theory, Equation (\ref{eq:Ie}) remains valid, but
the $\Phi$ operator, as well as the minimal impact parameter
$\rho_{\rm min}$ that segregates weak and strong collisions, change
with respect to the impact theory, and become
frequency-dependent. There is no rigorous way to make the transition
from one theory to the other. However, the profiles calculated
separately from each theory are almost identical in their common
overlapping region; the transition may thus be chosen quite
arbitrarily. Such a method is described in \citet{beauchamp97}.

\subsection{Summary of Approximations\label{sec:approx}}

The standard Stark broadening theory thus includes several
approximations, in addition to those already implied by the classical
path approximation. Ion dynamics is neglected, overlapping electronic
collisions are treated approximately, and the contribution from weak
and strong electronic collisions are modelled in distinct ways.

Another issue --- linked to the profile normalization process ---
occurs when one combines the standard theory with the one-electron
theory. As shown by \citet{Kolb58}, the standard theory yields, by
construction, a profile with a well-defined area equal to

\begin{equation}
     \int I(\omega) d\omega= \sum_{ab} |\langle a|\mathbf{d}|b\rangle|^2\ . 
\end{equation}

\noindent It is then trivial to normalize the profile to unit area as
required for synthetic spectrum calculations. When combining the
impact theory for the line core with the one-electron theory in the
wings, this normalization process is no longer valid, which
constitutes a formal weakness of this approach.

The purpose of the computer simulation approach is to lift these
approximations, while retaining the important aspects of the
semi-classical approach.

\section{Computer Simulations}\label{sec:simu}

\subsection{Overview of the Approach}\label{subsec:coco}

The purpose of the computer simulation approach is to integrate the
Schr\"odinger equation (Equation \ref{eq:Ua}) for the time evolution
operator $U(t,0)$ of the perturbed emitter, without resorting to the
truncated power series approximation of the analytical theory. This
method considers correctly the effect of overlapping collisions on
line broadening, regardless of their duration, thus unifying the
treatment of ions and electrons. The numerical integration process
treats indifferently weak and strong electronic collisions, whose
types were defined only in the context of the semi-analytical
theory. Moreover, the transition between the two electron broadening
regimes --- namely, the impact regime in the core and the one-electron
regime in the wings ---, as well as ion dynamics, are managed
naturally by this method.

The semi-classical approximation remains in force: the perturbers move
along classical paths, corresponding to straight lines when their
mutual interaction is neglected, and when the emitter is neutral. The
classical potential $V(t)$ is still approximated by the first terms in
the dipole expansion of the electrostatic potential. In this work, we
keep the first non-vanishing term that led to the Hamiltonian given in
Equation (\ref{eqn:H_a}). We apply the $\mu$-ion model from
\citet{Seidel82}, where a stationary emitter is located at the center
of a fixed simulation volume. In this model, it is the motion of the
perturbers in the rest frame of the emitter that is being simulated.

The thermal average of the autocorrelation function (Equation
\ref{eq:C7}), which corresponds to an average over the set of all
possible configurations of dynamic perturbers, is approximated by an
average over a large, but finite set of configurations, whose
statistical properties are representative of the infinite set. The
autocorrelation function is then written as an average over $N$
autocorrelation functions $C_i(t)$, each one calculated for a given
configuration $i$,
\begin{equation}
    C(t) = {1\over N} \sum_{i=1}^N C_i(t)\ .
\end{equation}
The line profile then follows from the numerical Fourier transform
(Equation~\ref{eq:stark9}). The whole process guarantees that the
profile can be properly normalized, thus resolving the issue raised in
Section \ref{sec:approx}.

Each dynamic configuration consists of a spherical simulation volume
of finite radius $R$, a single emitter located at the center of the
sphere, and a set of charged perturbers moving along classical paths
and generating a time-varying electric field ${\bf F}(t)$ at the
position of the emitter. The radius must ensure that the expected
statistical distribution of the electric field intensity is
recovered. As discussed by \citet{H88}, this radius must be of the
order of the Debye radius.

The value of $C_i$ --- the autocorrelation function of the $i$th
scenario --- is calculated for a series of discrete times $t_k =
k\Delta$, in two separate steps. First, a time sequence of electric
field ${\bf F}_i(t_k)$ at the origin of the simulation volume is
generated from the spatial distribution of moving perturbers at that
time. Equation (\ref{eq:Ua}) is then integrated in order to get the
time evolution operator $U_i(t_k)$. The validity of the simulation
method thus rests on the procedure for generating the dynamic
configurations, and for calculating the time evolution operator.

\subsection{Properties of the Simulation Volume}\label{subsec:gen}

\subsubsection{The Perturber-Emitter Interaction}\label{subsec:eleccons}

In these computer simulations, perturbers follow straight line
trajectories, and the effect of the spatial correlation produced by
their mutual interaction is thus neglected. As discussed in Section
\ref{sec:principles}, this correlation has a measurable impact on the
statistical distribution $W(F)$ of the microfield. A more realistic
simulation should take into account the effect on their trajectory of
the Coulomb interaction between the perturbers. The resulting
distribution of the microfields would then be in agreement with the
Hooper distribution, but at the cost of a prohibitive computing time.

The microfield distribution resulting from computer simulations with
rectilinear trajectories corresponds to the Holtsmark distribution, a
special case of the $W(F)$ distribution for independent particles. In
such calculations, the perturber-emitter interaction is described by a
Coulomb field. A way to recover the proper microfield distributions
with rectilinear trajectories is to replace the physically valid
Coulomb potential with a Debye potential
\citep{Gig03,Stambulchik2007}. The microfield distribution is then
well reproduced, including in the far wings where the contribution
from individual strong collisions become important. In what follows,
we adopt a Debye potential to describe the required perturber-emitter
interaction.

\subsubsection{The Simulation Volume and the Impact Coordinate System}\label{subsect:simuvol}

The simulation volume consists of three concentric spheres, all
centered on the position of the emitter. The smallest sphere
corresponds to the exclusion volume, with a radius of the order of the
Bohr radius. The particles never cross this volume in order to avoid
both penetrating and high-energy collisions, for which case the
classical path approximation breaks down. The second sphere, called
the calculation volume, includes the perturbers that contribute to the
electric field at the position of the emitter. In our simulations, its
radius is set to three times the Debye radius computed by considering
all perturbers (electrons and ions). \citet{H88} demonstrated that the
resulting microfield distribution reproduces adequately the
theoretical distribution $W(F)$ from \citet{Hoop68}, discussed in
Section \ref{subsec:eleccons}, if the perturber-emitter interaction is
described by a Debye potential. The third and largest sphere, called
the simulation volume, is limited by a radius equal to three times the
Debye radius of the electrons (excluding ions). Among the perturbers
in the simulation volume, only those present in the calculation volume
contribute to the electric field, allowing the possibility of
simulating the variation of the local number density in the
calculation volume. This important aspect was first introduced by
\citet{H88}. The number of particles in the calculation volume then
follows a Poisson distribution.

The coordinate system that gives the best control over the trajectory
generation process is the impact coordinate system, described by
\citet{H88}. These coordinates are :
\begin{enumerate}
    \item the impact parameter $b$, which is the distance of closest approach between the emitter and the perturber,
    \item the velocity $v$ of the perturber,
    \item the time $t_0$ of closest approach,
    \item the angle $\alpha$ that defines the trajectory of the perturber in the plane formed by the point of closest approach $\mathbf{b}$ and the velocity vector,
    \item the angles $\phi$ and $\theta$ that define the orientation of the velocity vector.
\end{enumerate}
In this coordinate system, the position of the perturber is 
\begin{equation}
\label{eq:distradimp2article}
    {\bf r}(t)=b(\hat{v_1}\cos\alpha+\hat{v_2}\sin\alpha)+\hat{v_3}v(t-t_0)\,,
\end{equation}
\noindent
where the vectors $\hat{v_1}$, $\hat{v_2}$, and $\hat{v_3}$ form an orthonormal basis, 
\begin{equation}
\label{eq:vectorbasearticle}
\begin{aligned}
    \hat{v_1}= & (-\cos\phi\cos\theta\,,\,-\sin\phi\cos\theta\,,\,\sin\theta) \\
    \hat{v_2}= & (-\sin\phi\,,\,\cos\phi\,,\,0) \\
    \hat{v_3}= & (\cos\phi\sin\theta\,,\,\sin\phi\sin\theta\,,\,\cos\theta).
\end{aligned}
\end{equation}
In Equation (\ref{eq:distradimp2article}), ${\bf r}$ is perpendicular
to $\hat{v_3}$ when $t=t_0$, as required.

The advantages of this coordinate system are two-fold. First, an
appropriate choice of the coordinate $t_0$ constrains each perturber
to be inside the simulation volume at the initialization phase of the
simulation ($t = 0)$, once the velocity $v$ and the impact parameter
$b$ have been assigned. Second, a proper value of the impact parameter
further guarantees that the impact vector is within the simulation
volume, but outside the exclusion volume, in the vicinity of the
emitter.

\subsection{Generation and Reinjection of Particles}

One of the most challenging difficulties with computer simulations is
to maintain a consistent number of perturbers, in a fixed simulation
volume, which satisfies the joint distribution of space and velocity,
without introducing undesirable correlations. To deal with this
problem, \citet{Gig96} proposed a simulation approach that combines an
initialization phase and a particle reinjection process. Particles are
first placed inside the simulation volume at $t=0$. As the simulation
evolves, some particles leave the volume and are replaced by new
particles of the same type, which enter the volume and follow a
straight line trajectory inside the sphere. The outgoing particles are
excluded for the rest of the simulation.

The particle generation process is a crucial step in the
simulation. It must ensure that in a given simulation volume, the
statistical properties of the system are maintained at all times, and
at every single time considering all simulations running in
parallel. To do so, the coordinates of each perturber must be
generated randomly from their joint statistical distribution. Luckily,
this distribution can be written as a product of distributions of the
six phase space coordinates. A uniform random number generator can
then be used to generate a random number, which is then mapped onto
the desired value based on the statistical distribution of the
generated coordinates. Two mapping strategies are applied below: the
inverse transformation method \citep{Numrecipe}, and a method based on
cells \citep{Gig96}.

The statistical distribution of each coordinate is obtained with the
following reasoning. The probability that a perturber is found inside
an infinitesimal volume ${\bf d^3r\ d^3v}$ of the phase space is
proportional to the product of this volume (in order to satisfy the
constraint of an isotropic and homogeneous space) with a Boltzmann
weight $e^{-v^2/v_T^2}$. Using the Jacobian $|\partial ({\bf r},{\bf
  v}) / \partial (\alpha, t_0, b, \phi, \theta, v)|$, the joint
probability distribution in the impact coordinate system becomes
\begin{equation}
\label{eq:gogo1}
    f(\alpha, t_0, b, \phi, \theta, v) \propto bv^3e^{-v^2/v_T^2}\sin\theta\ .
\end{equation}
However, since the simulation volume has a finite radius, the
perturbers must also respect at $t =0$ the following condition

\begin{equation}
\label{eq:constr}
|{\bf r}|^2= b^2+v^2t_0^2 \le R^2\,,
\end{equation}
\noindent
which translates into a constraint on the $t_0$ coordinate
\begin{equation}
\label{eq:gogo3}
|t_0| \le t_{0, {\rm max}}\,,
\end{equation}
\noindent
where $t_{0, {\rm max}}$ is a function of both the velocity $v$ and the impact parameter $b$
\begin{equation}
\label{eq:t0maxarticle}
    t_{0,{\rm max}}\equiv \frac{\sqrt{R^2-b^2}}{v}\ .
\end{equation}
Hence, the variables $b$, $v$, and $t_0$ are not statistically
independent even if the distribution (Equation \ref{eq:gogo1}) is
separable. This statistical dependency is taken into account when the
initial coordinates of the particles are randomly generated.

The joint distribution of the three variables, written as
\begin{equation}
\label{eq:heger4article}
    f(t_0, b, v) \propto bv^3e^{-v^2/v_T^2}\ ,
\end{equation}
can be expressed as the following product of conditional distributions
\begin{equation}
\label{eq:heger5article}
    f(t_0, b, v) = f_t(t_0|b, v)f_b(b|v)f_v(v)\ ,
\end{equation}
where the $|$ symbol stands for ``given''. Therefore, the 3-variable
$f$ distribution becomes the product of the distribution of $t_0$
(given $b$ and $v$), $b$ (given $v$), and $v$. Since $t_0$ is uniform
between the $\pm t_{0,{\rm max}}$ limits (for given $b$ and $v$), we
can write
\begin{equation}
\label{eq:gogo5}
f_t(t_0|b, v)={1\over 2t_{0, {\rm max}}}= {v\over 2\sqrt{R^2-b^2}} \ .
\end{equation}
By substituting this result into Equation (\ref{eq:heger4article}), we find that 
\begin{equation}
\label{eq:gogo6}
    f_b(b|v) \propto b \sqrt{R^2-b^2}\ ,
\end{equation}
an expression independent of $v$. This, in turn, implies that $b$ and
$v$ can be generated independently. Finally, the two previous results
can be combined to yield the Maxwell-Boltzmann distribution for the
velocity
\begin{equation}
\label{eq:gogo7}
    f_v(v)\propto v^2e^{-v^2/v_T^2}\ .
\end{equation}

The initialization process respects the joint statistical distribution
of the coordinates at time $t=0$. The values of the coordinates
$\alpha$, $\phi$, and $t_0$ are found by applying a linear
transformation on the output of the uniform number generator over the
interval (0,1), given that their statistical distribution is
uniform. The generation of the variable $\theta$ (with $f(\theta)
\propto \sin\theta$) is obtained with the inverse transformation
method.

The generation of the $b$ and $v$ coordinates follows a method
developed by \citet{Gig96} based on cells. The main advantage of this
method is to prevent a progressive displacement of the particles
towards the surface of the sphere, an apparent cooling of the plasma,
as well as an undesired correlation between the two coordinates. The
statistical distribution of $b$ is first divided into $N_p$ cells of
equal probability (where $N_p$ is the number of particles of type
$p$). Each particle is then assigned to a cell, and its $b$ value is
chosen randomly from within the interval associated with the cell. The
method is similar for $v$, but the statistical distribution is divided
into $N_p+1$ cells instead. Each particle is assigned to a $v$-cell,
leaving one empty cell. This empty cell plays a crucial role in the
reinjection process.

A particle reinjection process ensures that the joint statistical
distribution of coordinates remains unchanged (within normal
fluctuations) over the entire length of the simulation. When a
particle leaves the volume, a new particle of the same type is
randomly positioned at the surface of the simulation volume. Its
coordinates are then computed as follows. The new particle is assigned
to the same $b$-cell as the leaving particle, and the impact parameter
is generated randomly from within this cell. For the velocity, the
particle is assigned to the empty cell, and the cell of the leaving
particle becomes the empty cell. The value of the $t_0$ coordinate is
fixed, given $b$ and $v$, by the constraint that the particle is at
the surface of the outer sphere. Finally, the angular coordinates
($\alpha$, $\theta$, and $\phi$) are generated with the same method as
in the initialization phase.

\subsection{The Time Evolution Operator}\label{subsec:U}

Once an electric field sequence has been generated over a succession
of $N$ consecutive times $t_k \equiv k\Delta t$ (with $k$ = 0 to
$N-1$), the time evolution operator $U(t_k, 0)$ of the perturbed atom
must be calculated. The numerical integration of Equation (\ref{eq:U})
is performed using the solution for a Hamiltonian $H$ with no explicit
time dependence, $U(t_1, t_0) = e^{-i H(t_1-t_0)/\hbar}$, as well as
two properties of the time evolution operator: (1) the time continuity
condition, $U(t, t) = I$, and (2) the composition property,
$U(t_2,t_0)= U(t_2, t_1)\,U(t_1, t_0)$. These last two properties
allow us to write $U(t_k, 0)$ as a product of operators, computed over
all previous time intervals
\begin{equation}
\label{eq:SchrodU1article}
    U(t_k, 0)  = \prod_{j=0}^{k-1} U(t_{j+1}, t_{j})\ .
\end{equation}
This last equation is exact for any set of increasing values of $t_j$,
but is not applicable unless $U(t_{j+1}, t_j)$ can effectively be
calculated.

Following \citet{Gig96}, we make the approximation that the electric
field, and hence the Hamiltonian operator $H$, remain constant in
every time interval $(t_j, t_{j+1})$. These are denoted by ${\bf F}_j$
and $H_j$, respectively. The following Schr\"odinger equation
\begin{equation}
\label{eq:schrodUtk}
    i\hbar\,\frac{dU(t, t_{j-1})}{dt}=[H_0+e\mathbf{F}_{j-1}\cdot\mathbf{R}]\,U(t, t_{j-1})
\end{equation}
is then solved, one evolution operator at a time, 
\begin{equation}
\label{eq:Udyn1}
    U(t_j,t_{\mathrm{j-1}})=e^{-iH_{j-1}\Delta t/\hbar}\ ,
\end{equation}
which leads to
\begin{equation}
\label{eq:SchrodU5}
    U(t_k, 0)  = \prod_{j=1}^{k} e^{-iH_{j-1}\Delta t/\hbar} = e^{-iH_{k-1}\Delta t/\hbar}\ U(t_{k-1}, 0)\ .
\end{equation}
This numerical integration scheme guarantees that the time evolution operator is unitary.

The matrix elements of the time evolution operator are defined in the
base $|a\ket$ of the unperturbed states of the helium atom in Equation
(\ref{eq:C7}). It is however much simpler to first diagonalize $U(t_k,
t_{k-1})$ in the base of the perturbed states $|\alpha\ket$, specific
to this time interval, and then to carry out the unitary
transformation that transforms $U$ into the base of the unperturbed
states. As discussed in the Appendix \ref{sec:per}, the
diagonalization of the operator $H$ with a constant electric field
yields the energies $E_{\alpha}$ of the perturbed states, and the
transformation matrix with coefficients $\bra a | \alpha\ket $. This
process is repeated at every time step for the operator $H_{k-1}$. The
matrix elements of $U$ in the unperturbed basis then become
\begin{equation}
\label{eq:SchrodU6}
\begin{aligned} 
    \bra a|U(t_k, 0)|a'\ket    
    &= \sum_{\alpha \alpha' a''}
    \bra a|\alpha\ket 
    \bra \alpha|e^{-iH_{k-1}\Delta t/\hbar}|\alpha'\ket 
    \bra \alpha' | a''\ket \ 
    \bra a''| U(t_{k-1}, 0)|a'\ket\\
    &= \sum_{\alpha a''}\bra a|\alpha\ket e^{-iE_{\alpha}\Delta t/\hbar} \bra \alpha | a''\ket \ 
    \bra a''| U(t_{k-1}, 0)|a'\ket \ ,
\end{aligned}
\end{equation}
where the subscript $k-1$ is omitted for the perturbed energies
$E_\alpha$, and perturbed states $|\alpha\ket$ and $|\alpha'\ket$.

This integration procedure assumes that the variation of the electric
field is negligible during the time step $\Delta t$. The latter must
thus be chosen carefully to insure that this approximation remains
valid, while keeping computing time to a minimum. A characteristic
time for the temporal variation of the electric field is the collision
time of the fastest perturbers, i.e.~the electrons. We use the time
step proposed by \citet{Gig96}, which is one hundredth of the electron
collision time
\begin{equation}
\label{eq:deltt}
    \Delta t= 0.01\frac{r_0}{v_T}= 0.01\bigg(\frac{3}{4\pi N_e}\bigg)^{1/3}\bigg(\frac{2kT}{\mu_e}\bigg)^{-1/2}\ ,
\end{equation}
\noindent
where $r_0$ is the typical distance between plasma charges whose
electron number density is $N_e$, $v_T$ is the thermal velocity of the
electrons, $k$ is the Boltzmann constant, and $T$ is the plasma
temperature. The thermal velocity is computed with the reduced mass
$\mu_e$, consistent with the $\mu$-ion model.

Most spectral lines presented in this work were computed by averaging
the autocorrelation function over 25,000 simulation volumes, each one
producing a time sequence of 50,000 electric field vectors ${\bf
  F}(t_k)$. The energy levels $E_\alpha$ and states $|\alpha\ket$ of
the perturbed helium atom were calculated at each time step $t_k$, in
order to update the time evolution operator and the autocorrelation
function $C(t_k)$. This calculation was parallelized on a cluster of
computers. The typical computing time for a line at a given electronic
density and temperature varied from 4095 to 32,760 CPU hours.

\section{Results}\label{sec:comp}

We produced detailed profiles for six neutral helium lines, only two
of which have previously published profiles obtained with the computer
simulation approach, to our knowledge. The corresponding transitions
show a wide range of behaviors, which allow us to test our own
computer simulations under various physical conditions. Note that our
line profiles are not convolved with a Doppler profile, to allow a
direct comparison with those of \citet{Gig09} and \citet{Lara12}. We
also compare our results with a new series of line profiles similar to
those described by \citet{beauchamp97}, based on the standard Stark
broadening theory. These profiles have been calculated explicitly for
this work, using the same approximations as in our computer
simulations and neglecting the correction for the occupation
probability formalism introduced by \citet{Mihalas88}. More
specifically (1) the contribution of the lower levels to the
broadening is neglected, (2) the perturbed upper levels are restricted
to those with the same quantum number $n$ as the upper level of the
permitted transition, (3) the dipole matrix elements are calculated
with the method of \citet{Shomo68}, and (4) the integral over the
ionic field (Equation \ref{eqn:CtStandard}) extends to infinity; it is
not truncated to the critical value at which the highest Stark levels
with a given $n$ merge with other levels
\citep{Mihalas88,Seaton90,beauchamp97}. In addition, the two sets of
profiles were generated for the same grid of temperatures, electron
densities, and frequency, allowing for a more direct comparison. Since
ions are assumed to be quasi-static in the standard Stark broadening
theory, differences between both sets of calculations are expected
particularly in the line cores, where ion dynamics --- neglected in
the standard theory --- is properly managed by the computer simulation
approach \citep{Gig09}.

In what follows, we present our profile calculations for the following
helium lines: He {\sc i} $\lambda$4471 $2P^3-4D^3$, $\lambda$4922
$2P^1-4D^1$, $\lambda5877$ $2P^3-3D^3$, $\lambda6678$ $2P^1-3D^1$,
$\lambda4026$ $2P^3-5D^3$, and $\lambda4144$ $2P^1-6D^1$. The He {\sc
  i} $\lambda\lambda$4471 and 4922 lines are the strongest in DB white
dwarf spectra, and they represent special cases for which different
implementations of the Stark broadening theory have been compared over
the last fifty years or so
\citep{Griem68,BCS69,Calisti88,Schoning94,beauchamp97,Gig09,Lara12}. Our
grid covers electron number densities from $N_e=1\times10^{14}$
cm$^{-3}$ to $6\times10^{17}$ cm$^{-3}$, and temperatures of
$T=10,000$~K, $20,000$~K, and $40,000$~K, which are relevant for
synthetic spectrum calculations. We are interested here in lower
densities in order to reproduce the physical conditions encountered in
the upper atmospheric layers of white dwarfs where the line cores are
formed. We also present exploratory calculations at $N_e=10^{16}$
cm$^{-3}$ and $T=20,000$~K for four additional helium lines: He {\sc
  i} $\lambda\lambda5877$ and 6678, whose upper level has the
principal quantum number $n = 3$, as well as He {\sc i}
$\lambda\lambda4026$ and $4144$, with $n$ higher than 4.

\subsection{Helium Transitions with Upper Level $n=4$}

\subsubsection{He {\sc i} $\lambda 4471$}

\begin{figure}
    \centering
    \includegraphics[width=16.5cm]{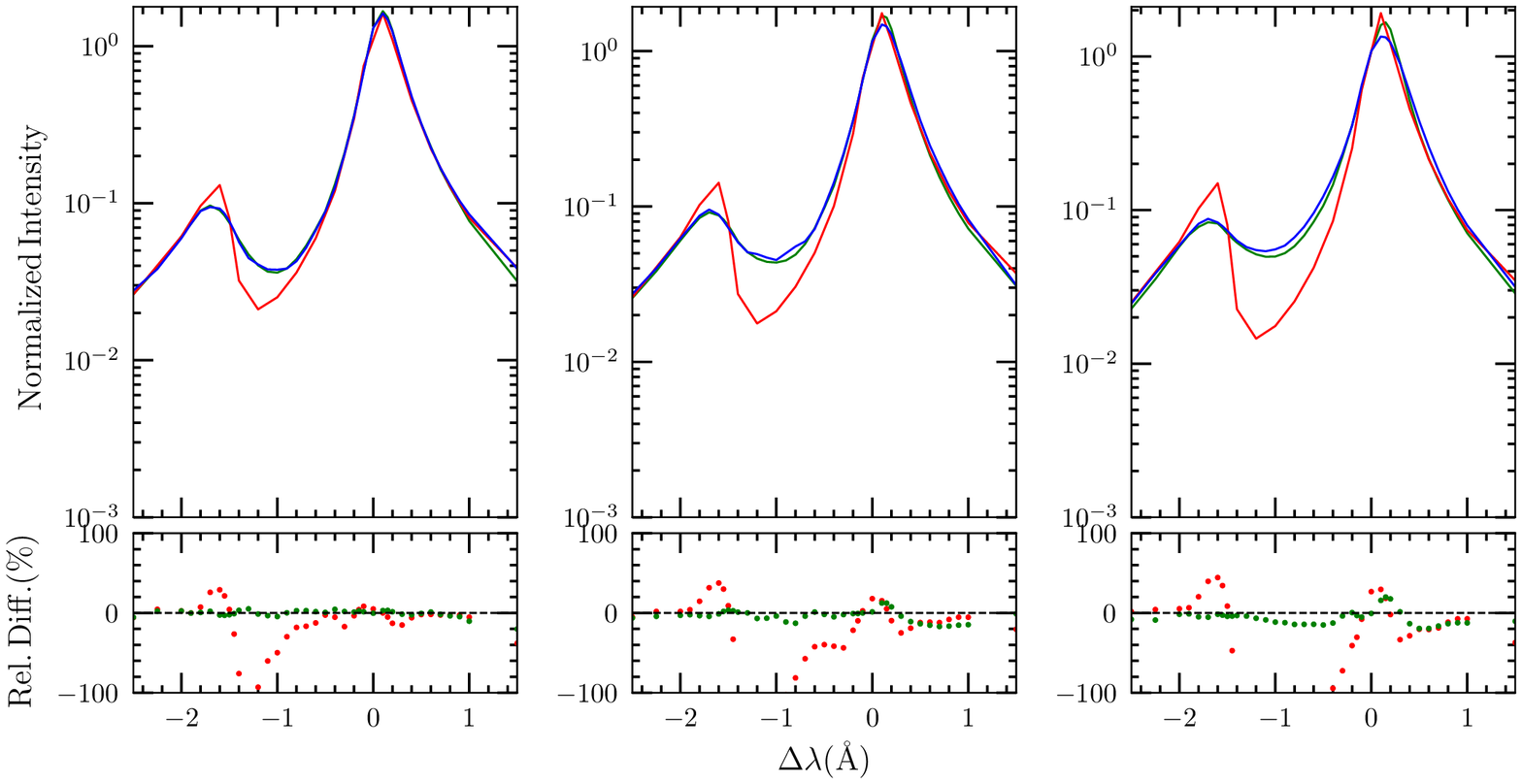}
    \caption[He {\sc i} $\lambda$4471 line comparison for
      $N_e=10^{15}$ cm$^{-3}$ ]{He {\sc i} $\lambda$4471 line profiles
      obtained from our computer simulations (blue line) compared with
      those of \citet[][red line]{beauchamp97} and \citet[][green
        line]{Gig09}. The profiles are shown for $N_e=10^{15}$
      cm$^{-3}$ and for temperatures of $T=10,000$~K, $20,000$~K, and
      $40,000$~K, from left to right. For a better comparison, the
      relative differences (in \%) are displayed below each panel with
      the same colors as those used in the upper panel.}
    \label{fig:p44711e15}
\end{figure}

\begin{figure}
    \centering
    \includegraphics[width=16.5cm]{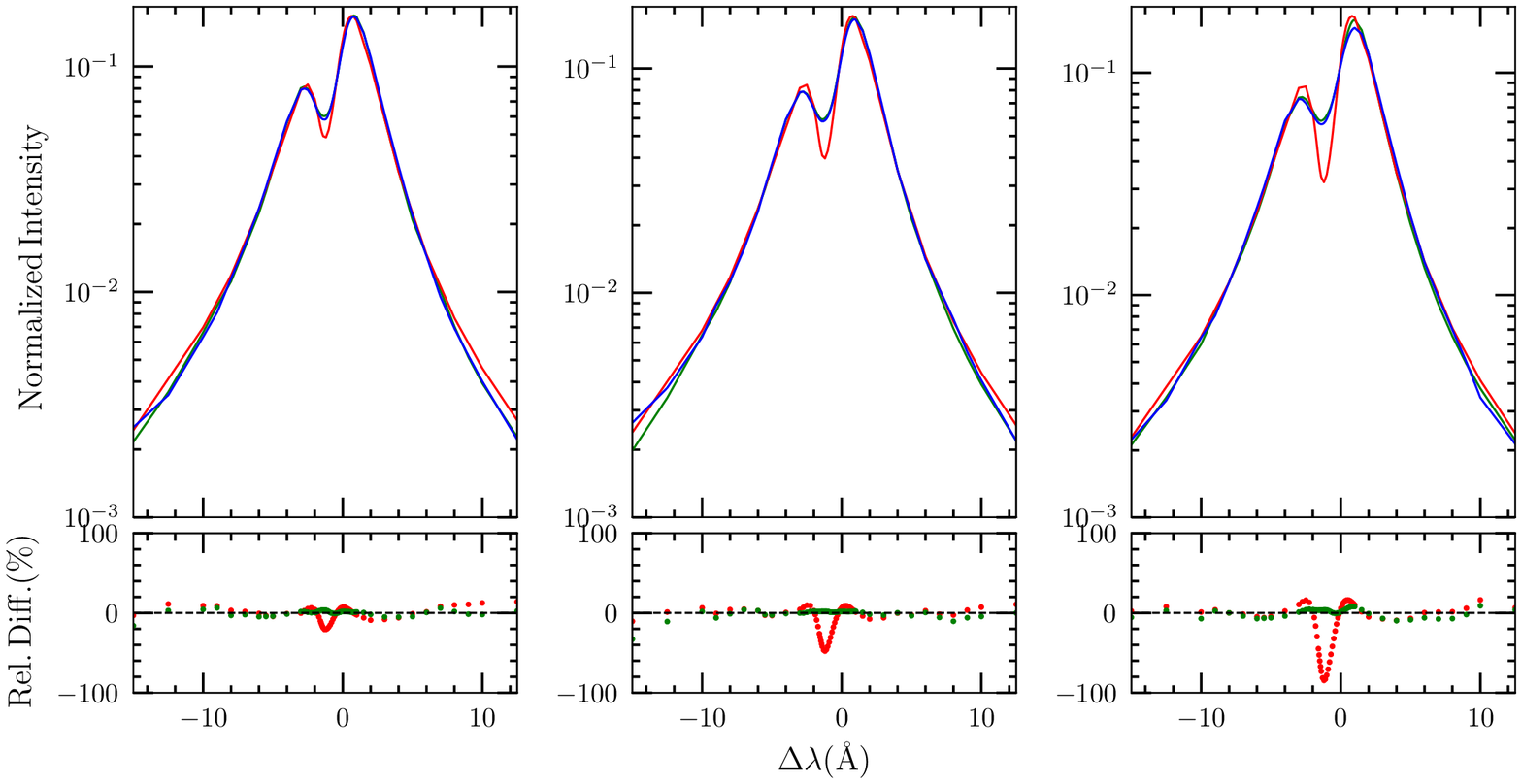}
    \caption[He {\sc i} $\lambda$4471 line comparison for an electron
      number density of $10^{16}$ cm$^{-3}$ ]{Same as Figure
      \ref{fig:p44711e15} but with $N_e=10^{16}$ cm$^{-3}$.}
    \label{fig:p44711e16}
\end{figure}

\begin{figure}
    \centering
    \includegraphics[width=16.5cm]{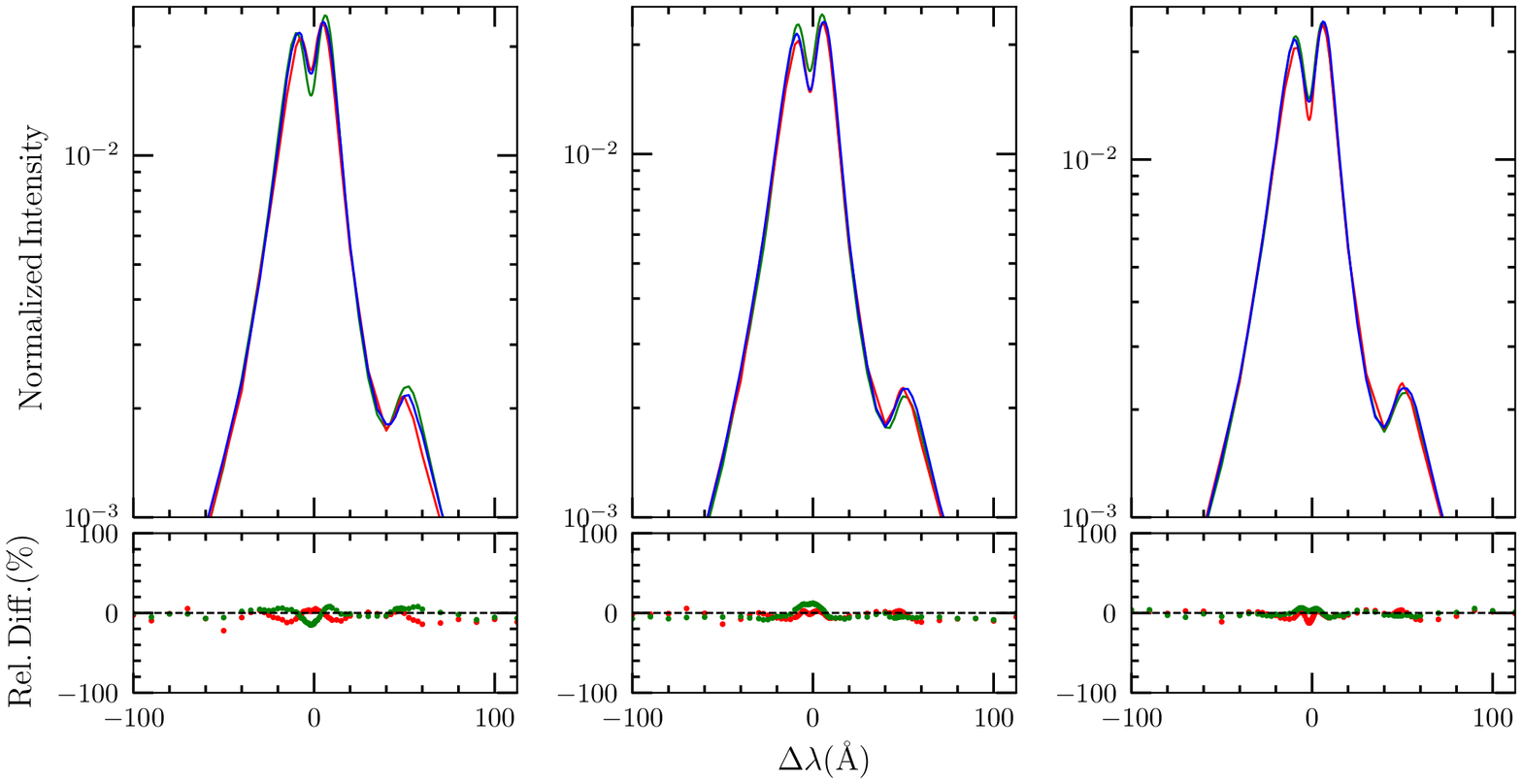}
    \caption[He {\sc i} $\lambda$4471 line comparison for an electron
      number density of $10^{17}$ cm$^{-3}$ ]{Same as Figure
      \ref{fig:p44711e15} but with $N_e=10^{17}$ cm$^{-3}$.}
    \label{fig:p44711e17}
\end{figure}

Profiles for the He {\sc i} $\lambda4471$ line are displayed in
Figures \ref{fig:p44711e15} to \ref{fig:p44711e17} for various values
of temperature and electron density. The profiles for $N_e=10^{15}$
cm$^{-3}$, shown in Figure \ref{fig:p44711e15}, correspond to the
lowest density available in the calculations of \citet{Gig09}. Any
difference between our computer simulations and theirs will require
some explanation, because we followed essentially the same approach.

As expected, the main difference between our profiles, obtained from
computer simulations, and those of \citet{beauchamp97}, based on the
standard Stark broadening theory, is for the low-density regime, near
the line core of the permitted and forbidden components, as well as
between them. This difference, of the order of a factor of 2 for some
profiles, is due to ion dynamics, whose effect is known to be
important at low densities \citep{BCS74,Gig96,Gig09,Ferri14}. Both
sets of profiles become almost identical at high densities, however,
when ion dynamics becomes negligible. We conclude that the main
difference between the two sets of profiles is in the treatment of ion
dynamics, and that the other approximation made in the work of
\citet{beauchamp97}, namely the treatment of the transition between
the impact and the one-electron regimes, seems appropriate at high
densities, at least for this particular line.

\begin{figure}
    \centering
    \includegraphics[width=16.5cm]{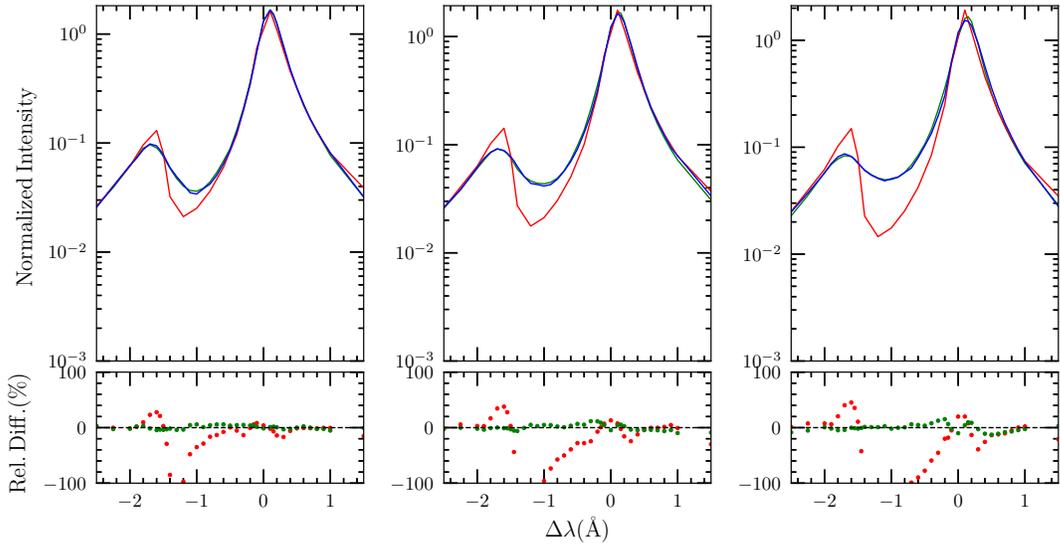}
    \caption[He {\sc i} $\lambda$4471 line comparison for an electron
      number density of $10^{15}$ cm$^{-3}$ without a local density
      variation]{Same as Figure \ref{fig:p44711e15}, but without
      taking into account local density variations, as properly
      implemented in the simulations of \citet{H88}. Our profiles
      (blue line) then resemble more those of \citet[][green
        line]{Gig09}, where this effect was neglected.}
    \label{fig:p44711e15nHeg}
\end{figure}

Another reason must be invoked to account for the differences between
our profiles and those of \citet{Gig09}, given that our simulation
environment is strongly inspired by their work. In this case, the main
difference between our approach and theirs lies in the design of the
simulation volume. As described in Section \ref{subsect:simuvol}, our
model consists of a calculation volume, defined as the volume that
includes the perturbers contributing to the electric field at the
position of the emitter, as well as a larger simulation volume. The
particle reinjection process involves the simulation volume, which
allows for local density variations, as properly implemented in the
computer simulations of \citet{H88}. In the work of Gigosos \&
González, these two spheres are identical, and the number of
perturbers therefore remains constant in the calculation volume.  No
study has shown that one approach is better than the other. For
comparison purposes, we generated a new grid of profiles with
identical calculation and simulation volumes, the results of which are
displayed in Figure \ref{fig:p44711e15nHeg}. Our profiles and those of
Gigosos \& González now overlap almost perfectly. The comparison
between the results shown in Figures \ref{fig:p44711e15} and
\ref{fig:p44711e15nHeg} reveals that the impact of local density
variations is important mostly in the region between the permitted and
the forbidden components ($\Delta \lambda\sim -1$ \AA), as well as in
the core of the permitted component. The minor differences that remain
at higher temperature in the core of the latter component (see
rightmost panel of Figure \ref{fig:p44711e15nHeg}) still require an
explanation.

In general, the three grids of profiles for He {\sc i} $\lambda$4471
are in excellent agreement at high densities, where ion dynamics is
less important, thus demonstrating the validity of the semi-analytical
approach in this regime.

\subsubsection{He {\sc i} $\lambda 4922$}

\begin{figure}
    \centering
    \includegraphics[width=16.5cm]{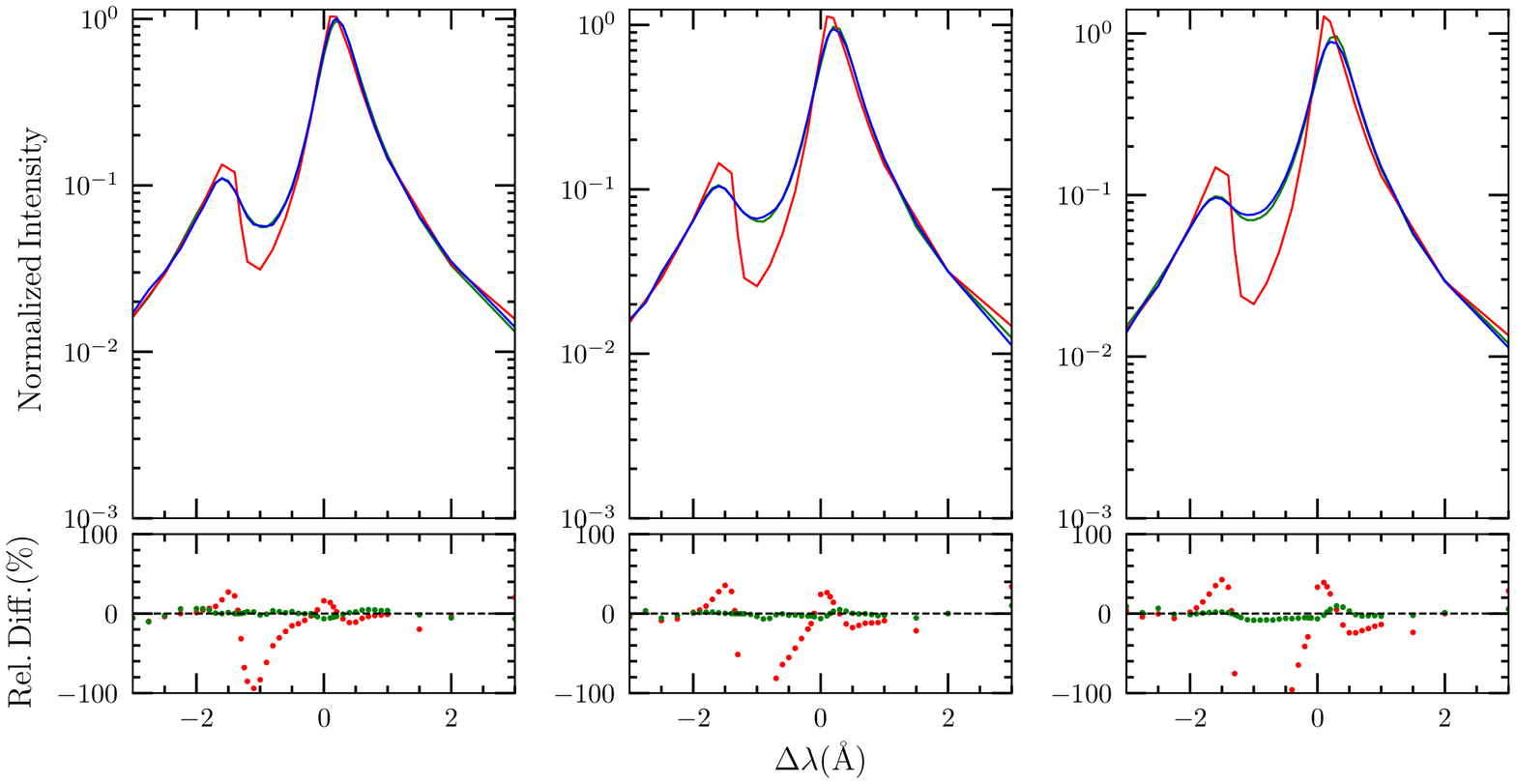}
    \caption[He {\sc i} $\lambda$4922 line comparison for an electron
      number density of $10^{15}$ cm$^{-3}$ ]{ He {\sc i}
      $\lambda$4922 line profiles obtained from our computer
      simulations (blue line) compared with those of \citet[][red
        line]{beauchamp97} and \citet[][green line]{Lara12}. The
      profiles are shown for $N_e=10^{15}$ cm$^{-3}$ and for
      temperatures of $T=10,000$~K, $20,000$~K, and $40,000$~K, from
      left to right. For a better comparison, the relative differences
      (in \%) are displayed below each panel with the same colors as
      those used in the upper panel.}
    \label{fig:p49221e15}
\end{figure}

\begin{figure}
    \centering
    \includegraphics[width=16.5cm]{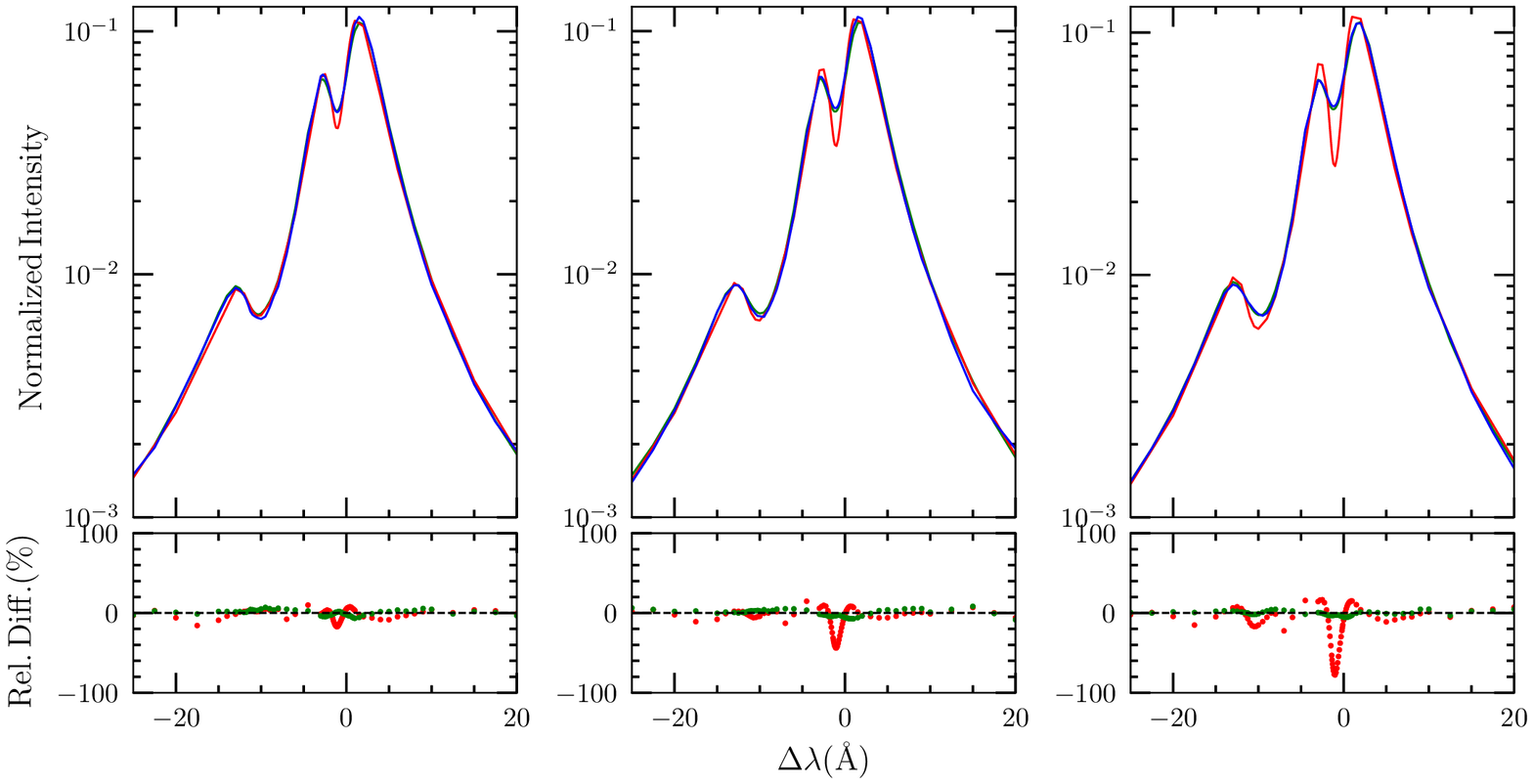}
    \caption[He {\sc i} $\lambda$4922 line comparison for an electron
      number density of $10^{16}$ cm$^{-3}$ ]{Same as Figure
      \ref{fig:p49221e15} but with $N_e=10^{16}$ cm$^{-3}$.}
    \label{fig:p49221e16}
\end{figure}

\begin{figure}
    \centering
    \includegraphics[width=16.5cm]{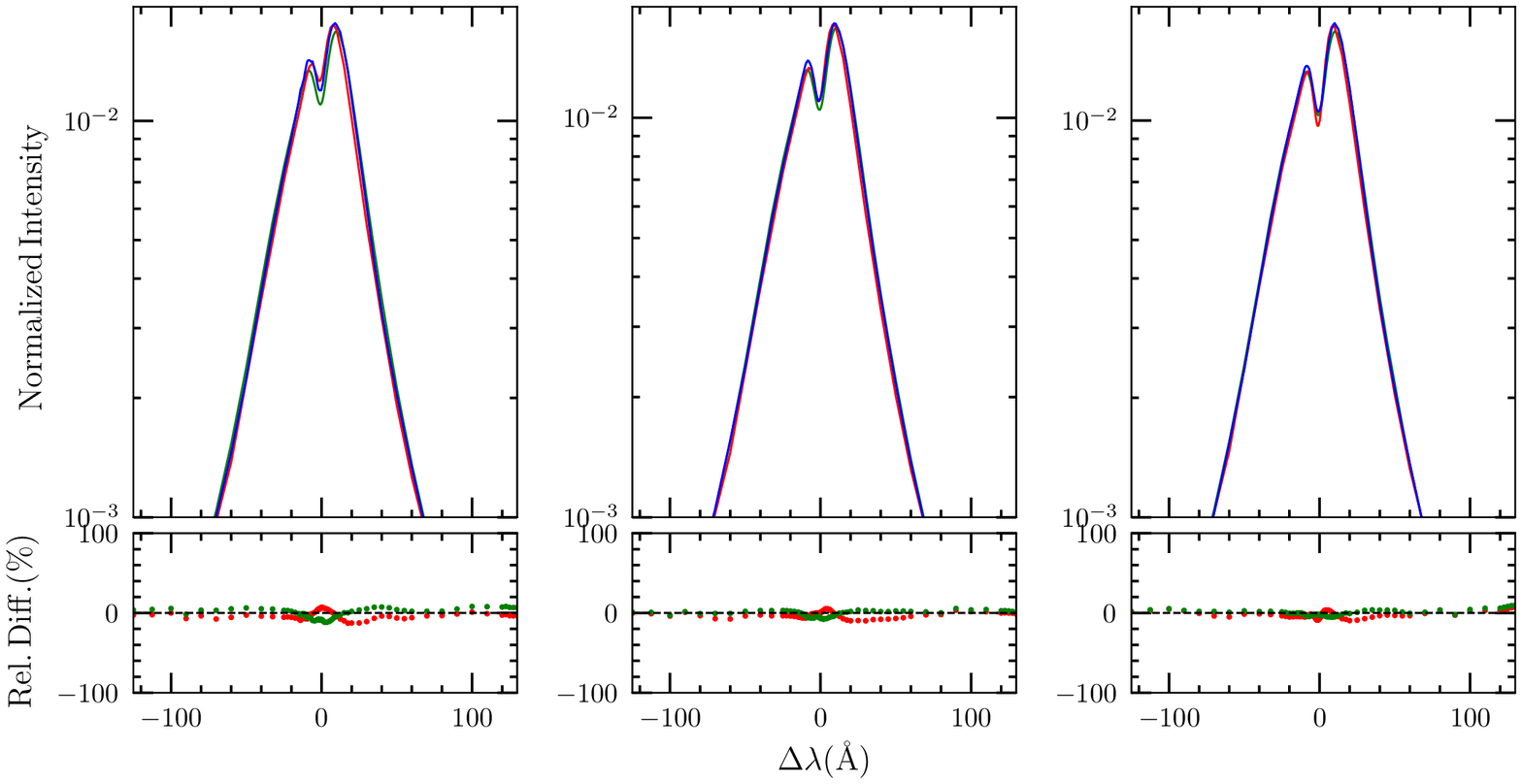}
    \caption[He {\sc i} $\lambda$4922 line comparison for an electron
      number density of $10^{17}$ cm$^{-3}$ ]{Same as Figure
      \ref{fig:p49221e15} but with $N_e=10^{17}$ cm$^{-3}$.}
    \label{fig:p49221e17}
\end{figure}

\begin{figure}
    \centering
    \includegraphics[width=16.5cm]{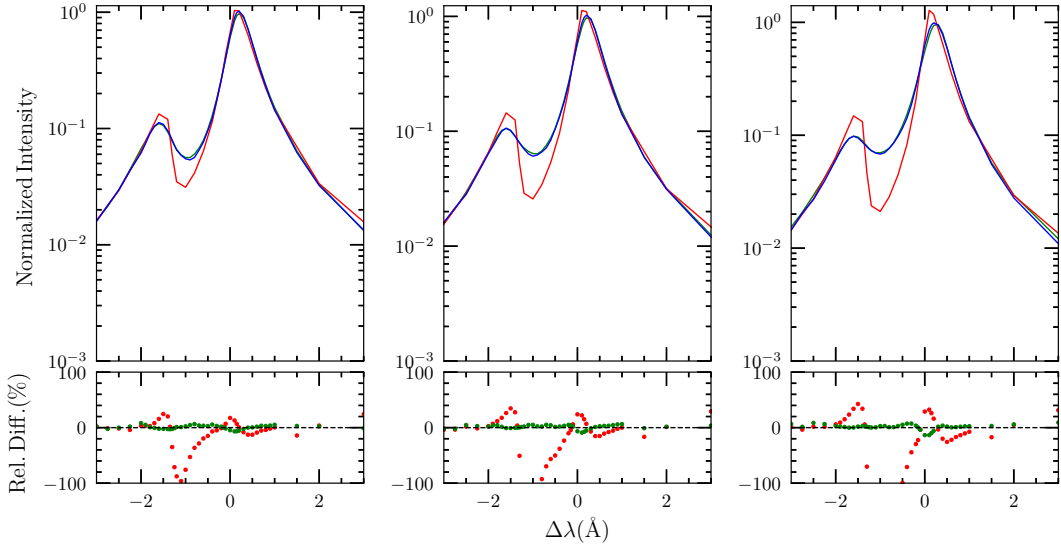}
    \caption[He {\sc i} $\lambda$4922 line comparison for an electron
      number density of $10^{15}$ cm$^{-3}$ without a local density
      variation]{ Same as Figure \ref{fig:p49221e15}, but without
      taking into account local density variations, as properly
      implemented in the simulations of \citet{H88}. The differences
      between our profiles (blue line) and those of \citet[][green
        line]{Lara12} are of the same order as those observed in
      Figure \ref{fig:p44711e15nHeg}.}
    \label{fig:p49221e15nHeg}
\end{figure}

The profiles for He {\sc i} $\lambda$4922 are presented in Figures
\ref{fig:p49221e15} to \ref{fig:p49221e17} for the same temperatures
and electron densities as before. The differences between our profiles
obtained from computer simulations and those calculated using the
standard Stark broadening can be explained again in terms of the
effect of ion dynamics, similar to that already discussed in the
context of the He {\sc i} $\lambda$4471 line, in particular at low
densities, near the line cores, and in the region between the
permitted component and its nearest forbidden component. The effect of
ion dynamics can also be observed near the forbidden component at
$\Delta\lambda \sim -10$ \AA\,\,in Figure \ref{fig:p49221e16}.  The
differences between our He {\sc i} $\lambda4922$ profiles and the
computer simulations from \citet[][see Figure
  \ref{fig:p49221e15}]{Lara12}, based on the same approach as
\citet{Gig09}, are smaller than the differences observed previously
for He {\sc i} $\lambda4471$, especially at $T=40,000$~K, although
they do occur in the same range of wavelength. As before, it is
possible to demonstrate that these differences are due to our
treatment of local density variations, by generating a new grid of
profiles where such variations are neglected. The results of this
experiment, displayed in Figure \ref{fig:p49221e15nHeg}, confirm this
interpretation. Hence, the origin of these differences, as well as
those for He {\sc i} $\lambda\lambda$4471 and 4922, is the inclusion
of local density variations in our computer simulations.

The results presented so far for He {\sc i} $\lambda\lambda$4471 and
4922 confirm that our computer simulation environment reproduces the
published profiles adequately, including those obtained from
independent simulations that take into account the contribution of
lower levels perturbation and interference to broadening. This also
demonstrates that the contribution of the lower states and
interference between the lower and higher states, neglected in our
simulations, are dominated by that of the upper states for these two
transitions. We expect that the relative contribution of the lower
states term will be more important for transitions from $n=2$ to $n=3$
(He {\sc i} $\lambda\lambda$5877 and 6678). This will be investigated
further in a future paper. Furthermore, we also validated the
versatility of our implementation by easily adapting our calculations
to different energy levels. We can now explore new He {\sc i}
transitions, characterized by upper levels with a principal quantum
number different than $n=4$, for which the only published detailed
profiles have been obtained from the standard Stark broadening theory.

\subsection{Helium Transitions with Upper Level $n = 3$}

\begin{figure}
    \centering
    \includegraphics[width=8.5cm]{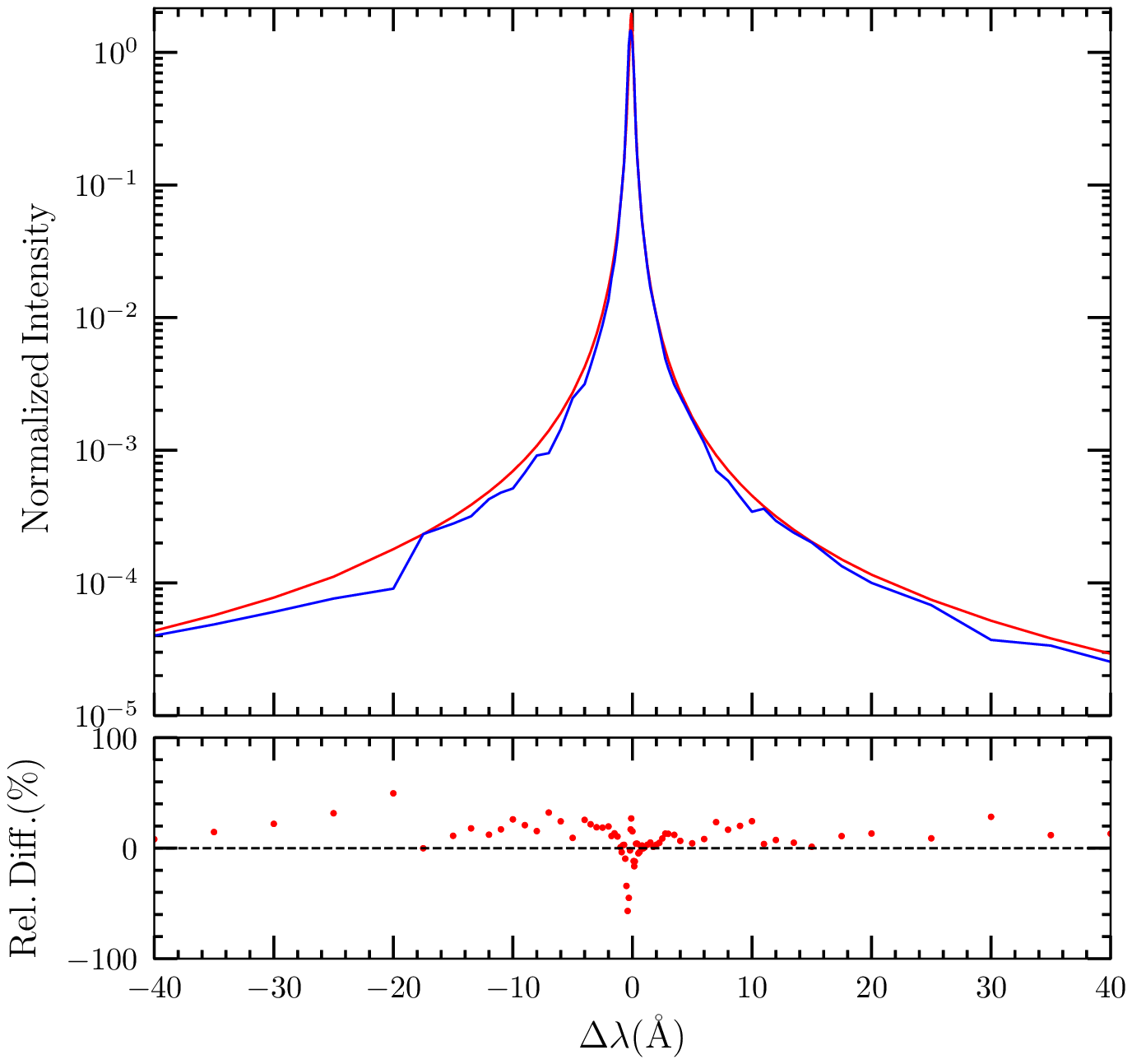}
    \caption[He {\sc i} $\lambda$5877 line comparison for an electron
      number density of $10^{16}$ cm$^{-3}$ ]{He {\sc i} $\lambda$5877
      line profile obtained from our computer simulations (blue line)
      compared with that of \citet[][red line]{beauchamp97}, for
      $N_e=10^{16}$ cm$^{-3}$ and $T=20,000$~K. The relative
      differences (in \%) are displayed at the bottom with the same
      colors as those used in the upper panel.}\label{fig:prof5877}
\end{figure}

\begin{figure}
    \centering
    \includegraphics[width=8.cm]{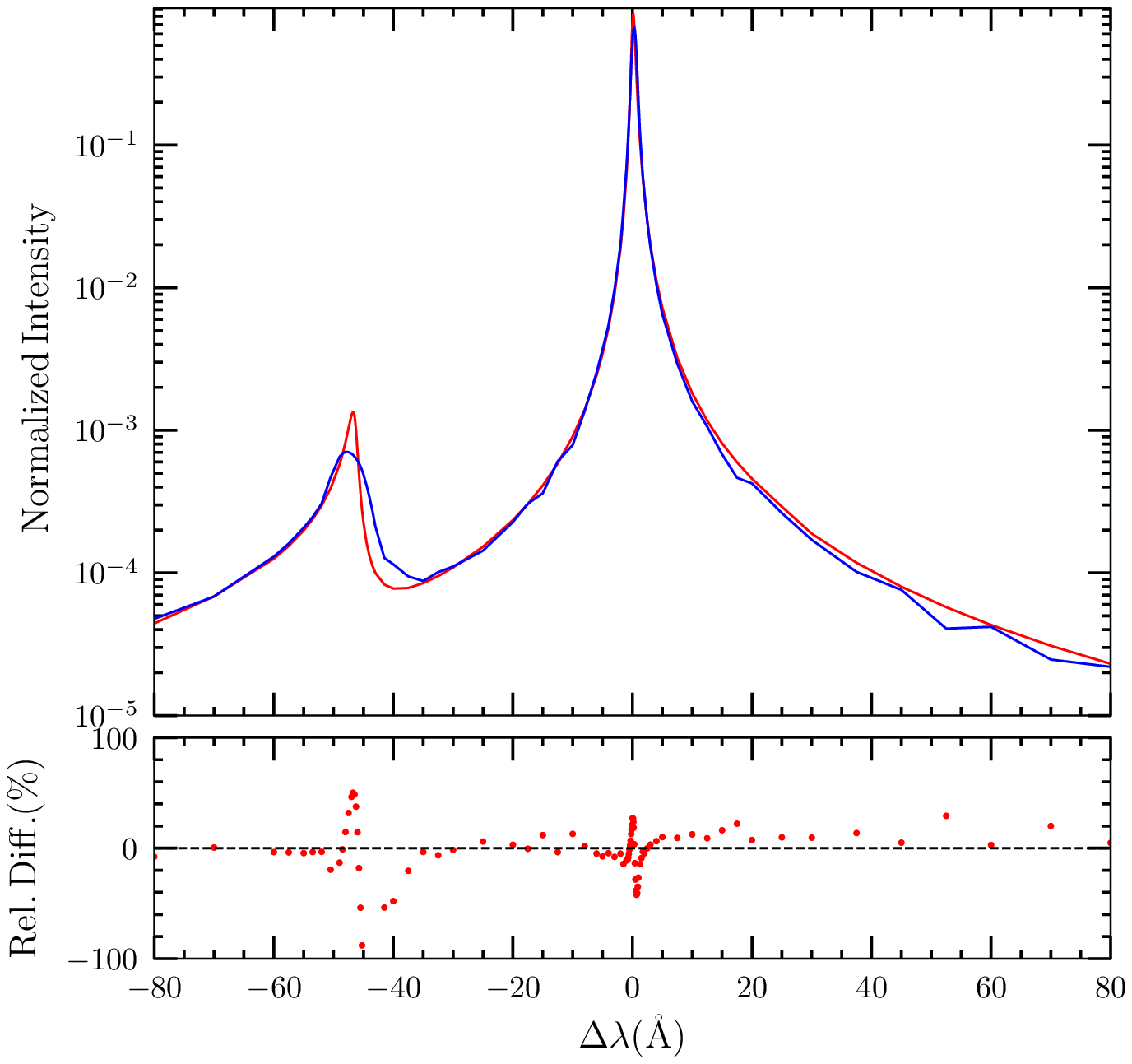}
    \caption[He {\sc i} $\lambda$6678 line comparison for an electron
      number density of $10^{16}$ cm$^{-3}$ ]{He {\sc i} $\lambda$6678
      line profile obtained from our computer simulations (blue line)
      compared with that of \citet[][red line]{beauchamp97}, for
      $N_e=10^{16}$ cm$^{-3}$ and $T=20,000$~K. The relative
      differences (in \%) are displayed at the bottom with the same
      colors as those used in the upper panel.}
    \label{fig:prof6678}
\end{figure}

\begin{figure}
    \centering
    \includegraphics[width=8.cm]{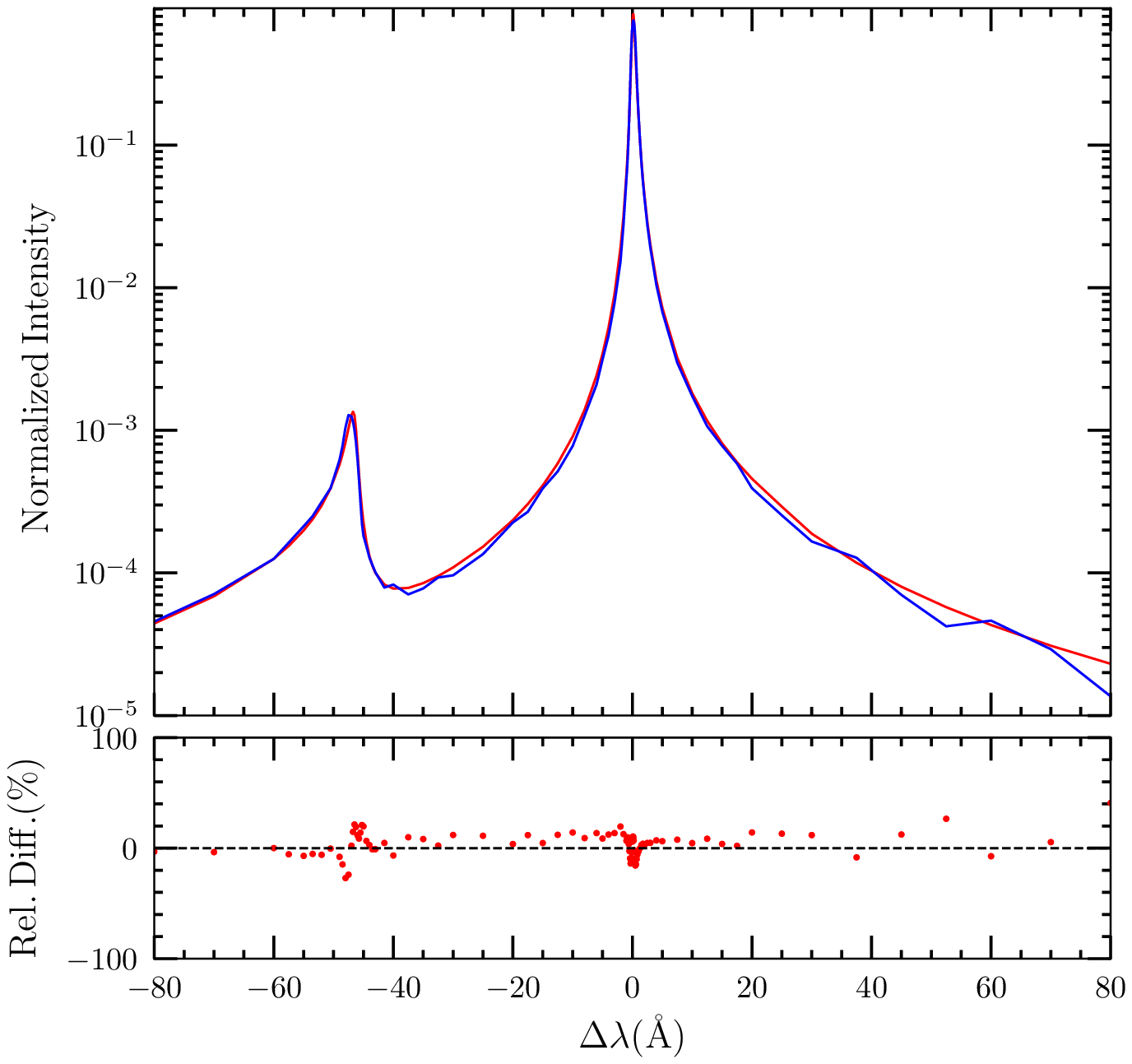}
    \caption[He {\sc i} $\lambda$6678 line comparison for an electron
      number density of $10^{16}$ cm$^{-3}$ with static ions]{Same as
      Figure \ref{fig:prof6678} but without taking ion dynamics into
      account in the computer simulation. The largest difference with
      the results displayed in Figure \ref{fig:prof6678} occurs for
      the forbidden component.}
    \label{fig:prof6678nions}
\end{figure}

We compare in Figures \ref{fig:prof5877} and \ref{fig:prof6678} the
profiles for the He {\sc i} $\lambda$5877 $2P^3-3D^3$ and
$\lambda$6678 $2P^1-3D^1$ lines, respectively, obtained from our
computer simulations with those calculated using the standard Stark
broadening theory \citep{beauchamp97}. The comparisons displayed here
differ significantly from those shown above for He {\sc i}
$\lambda\lambda$4471 and 4922. Neutral helium transitions involving an
upper state with a principal quantum number $n=3$ are generally
isolated lines at the electron densities considered in this work, and
they are characterized by a Lorentzian shape, with a weak and distant
forbidden component, resulting from the high energy differences
between the upper level of the permitted transition and its
neighboring levels. The closest forbidden component for He {\sc i}
$\lambda5877$ is $\lambda6068$ $2P^3-3P^3$. This forbidden component
constitutes a challenge for the simulation method, because its
extremely low intensity, especially in the wings, could easily be
misrepresented as a result of numerical noise. The profile for He {\sc
  i} $\lambda6678$ includes the very asymmetrical forbidden component
$\lambda6632$ $2P^1-3P^1$. Despite its apparent simplicity, a
Lorentzian-like line profile represents another challenge for computer
simulations. Indeed, since the permitted component of these two lines
is very narrow, and that the forbidden component is at much lower
intensities, it is necessary to improve the wavelength (or frequency)
resolution and to decrease the numerical noise. This implies
increasing both the number of time steps and simulation volumes. A
larger number of simulated volumes yields a better representation of
the microfield distribution for the stronger, less likely, electric
fields that contribute in the wings of the profile. For He {\sc i}
$\lambda$5877, a simulation time of 70,000 time steps and 50,000
simulation volumes were required to define properly the permitted
component and to reduce the numerical noise, respectively. Some
residual noise remains in the wings of the profile displayed in Figure
\ref{fig:prof5877}, despite these modifications. For He {\sc i}
$\lambda$6678, 60,000 time steps and 60,000 simulation volumes were
required to properly define the permitted and forbidden components.

The He {\sc i} $\lambda6678$ line profile represents an interesting
case. Although the shape of its permitted component is very similar
between both sets of calculations displayed in Figure
\ref{fig:prof6678}, there are important differences in the shape of
the forbidden component. The most plausible explanation for the more
rounded shape of the forbidden component obtained with our computer
simulations is the additional broadening produced by ion dynamics. To
test this hypothesis, we generated a profile with a version of our
computer simulation in which the ions are at rest. The results,
displayed in Figure \ref{fig:prof6678nions}, indicate that indeed, the
forbidden component calculated with quasi-static ions now closely
resemble that predicted by the standard Stark broadening theory
(despite the presence of numerical noise). The effect of ion dynamics
therefore remains important for this forbidden component, even at
densities as high as $N_e=10^{16}$ cm$^{-3}$.

So far, we have shown that our computer simulation framework
reproduces well the profiles of some isolated lines, as well as lines
with forbidden components that are well-separated from the permitted
component. We now consider the case of helium lines involving upper
levels for which $n > 4$, which are almost hydrogenic.

\subsection{Helium Transitions with Upper Level $n>4$}

\begin{figure}[h]
    \centering
    \includegraphics[width=8.cm]{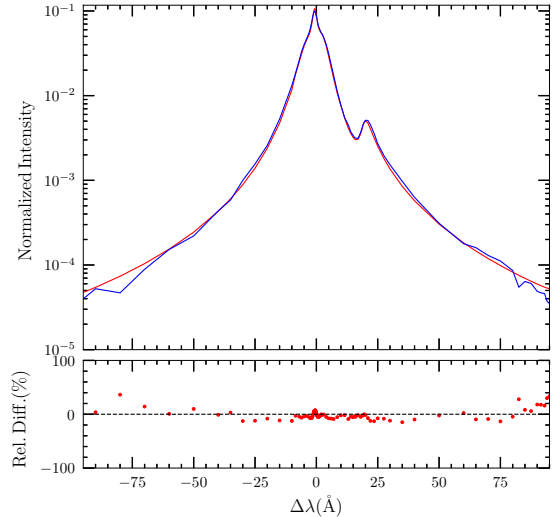}
    \caption[He {\sc i} $\lambda$4026 line comparison for an electron
      number density of $10^{16}$ cm$^{-3}$ ]{Same as Figures
      \ref{fig:prof5877} and \ref{fig:prof6678} but for the He {\sc i}
      $\lambda$4026 line. }
    \label{fig:p4026}
\end{figure}

\begin{figure}[h]
    \centering
    \includegraphics[width=8.cm]{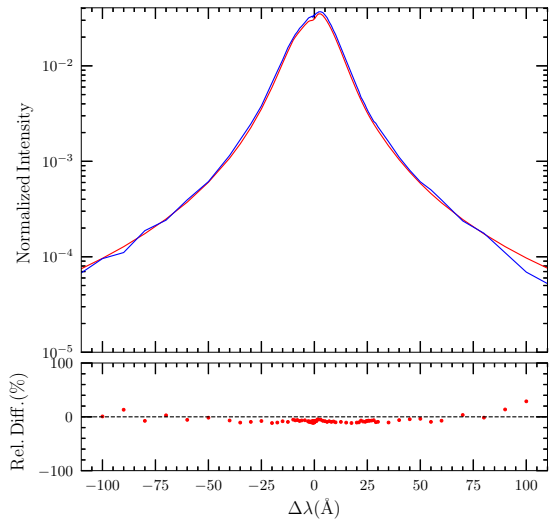}
    \caption[He {\sc i} $\lambda$4144 line comparison for an electron
      number density of $10^{16}$ cm$^{-3}$ ]{Same as Figures
      \ref{fig:prof5877} and \ref{fig:prof6678} but for the He {\sc i}
      $\lambda$4144 line. }
    \label{fig:p4144}
\end{figure}

Similar comparisons for the He {\sc i} $\lambda$4026 $2P^3-5D^3$ and
$\lambda$4144 $2P^1-6D^1$ lines are displayed in Figures
\ref{fig:p4026} and \ref{fig:p4144}, respectively. For these lines,
the permitted component is superimposed on the strong forbidden
components $2P-nP, F, G$ because these upper levels, of almost
identical energy in the isolated helium atom, are mixed by the
electric field. The width of the lines tends to increase with the
principal quantum number $n$ for a given electron density and
temperature. This behavior can be appreciated by comparing the results
shown in Figures \ref{fig:p4026} and \ref{fig:p4144} --- which both
correspond to transitions with an upper level $n=5$ and $n=6$,
respectively --- with the results displayed in Figures
\ref{fig:prof5877} and \ref{fig:prof6678} for $n=3$.

The calculation of these profiles is numerically demanding since the
dimension of the matrices for which the eigenvalues and eigenvectors
have to be extracted increases as $n^2$. However, the absence of
narrow structures in the profiles implies that a lower resolution in
wavelength (or frequency) is sufficient to represent the profile
correctly. The profile for He {\sc i} $\lambda4144$ ($n=6$) was
computed with 40,000 time steps, instead of the 50,000 time steps
required for the He {\sc i} $\lambda4026$ ($n=5$) profile. The latter
also shows the forbidden component $\lambda4045$ $2P^3-5P^3$, as well
as a complex behavior near the peak of the profile, where the
permitted component mixes with other forbidden components. The
profiles of these two lines calculated with our computer simulations
compare well with those produced with the standard Stark broadening
theory, at this electron density. Ion dynamics appears to have less
impact on these line profiles than for He {\sc i} $\lambda\lambda
4471$ and $4922$ at an electron density of $N_e=10^{16}$ cm$^{-3}$.

\subsection{Application to White Dwarf Model Spectra}

As discussed in the Introduction, the atmospheric parameters of white
dwarf stars can be measured using the so-called spectroscopic
technique, where absorption line profiles are compared with the
predictions of detailed model atmospheres. Given the limited number of
He {\sc i} profiles calculated in this paper, it is too early at this
stage to present a full-fledged spectroscopic analysis of DB
(helium-line) white dwarfs. However, it is still possible to perform a
few quantitative comparisons with preexisting material.

The calculations of model atmospheres require to solve simultaneously
the radiative transfer equation, the hydrostatic equilibrium equation,
and the radiative equilibrium equation, which is replaced by a flux
conservation equation in the presence of convective energy
transport. It is customary under most white dwarf conditions to assume
local thermodynamic equilibrium (LTE), which allows the use of the
standard Saha-Boltzmann equations to describe the ionization and
excitation equilibrium. The solution of these equations provides the
temperature and pressure structure as a function of optical depth. In
a second step, the detailed emergent monochromatic flux --- referred
to as the synthetic spectrum --- can be calculated by solving the
radiative transfer equation using the previously calculated
thermodynamic structure. The most important physical quantity in these
calculations is the extinction (or opacity) coefficient, $\chi_\nu$,
which describes all absorption (bound-bound, bound-free, free-free)
and diffusion (Thomson and Rayleigh scattering) processes. Thus in
principle, the line absorption profiles (bound-bound opacity) may
affect the emergent flux, but also the overall atmospheric
structure. Since we have only a limited number of improved He {\sc i}
profiles at our disposal, we will compare, in what follows, the
results of synthetic spectra using two different sets of line profile
calculations, but by keeping the thermodynamic structures fixed.

Our model atmospheres are similar to those employed by
\citet{Bergeron2011}. These are built from the model atmosphere code
described at length in \citet{Tremblay2009} and references therein,
which incorporate the Stark profiles of neutral helium of
\citet{beauchamp97}. As discussed above, these detailed profiles for
over 20 lines of neutral helium take into account the transition from
quadratic to linear Stark broadening, the transition from the impact
to the quasistatic regime for electrons, as well as forbidden
components. The input parameters of model atmospheres are the
effective temperature ($\Te$), the surface gravity ($\logg$), and the
chemical composition, which we assume here to be pure helium (H/He=0).

The comparison of synthetic spectra calculated using the profiles of
\citet{beauchamp97} with those obtained from our computer simulations
differ from the comparisons already discussed in the previous
subsections. Indeed, a stellar line profile is not representative of
unique temperature and electron density, but instead is formed at
different depths in the stellar atmosphere. For instance, the line
core, where the opacity is the largest, is formed high in the
atmosphere where both the temperature and density are low. In
contrast, the line wings, where the opacity is significantly reduced,
are formed in the deeper photospheric regions, where the temperature
and density are much larger, and more characteristic of the regions
where the continuum is formed.

\begin{figure}
    \centering
    \includegraphics[width=16.5cm ]{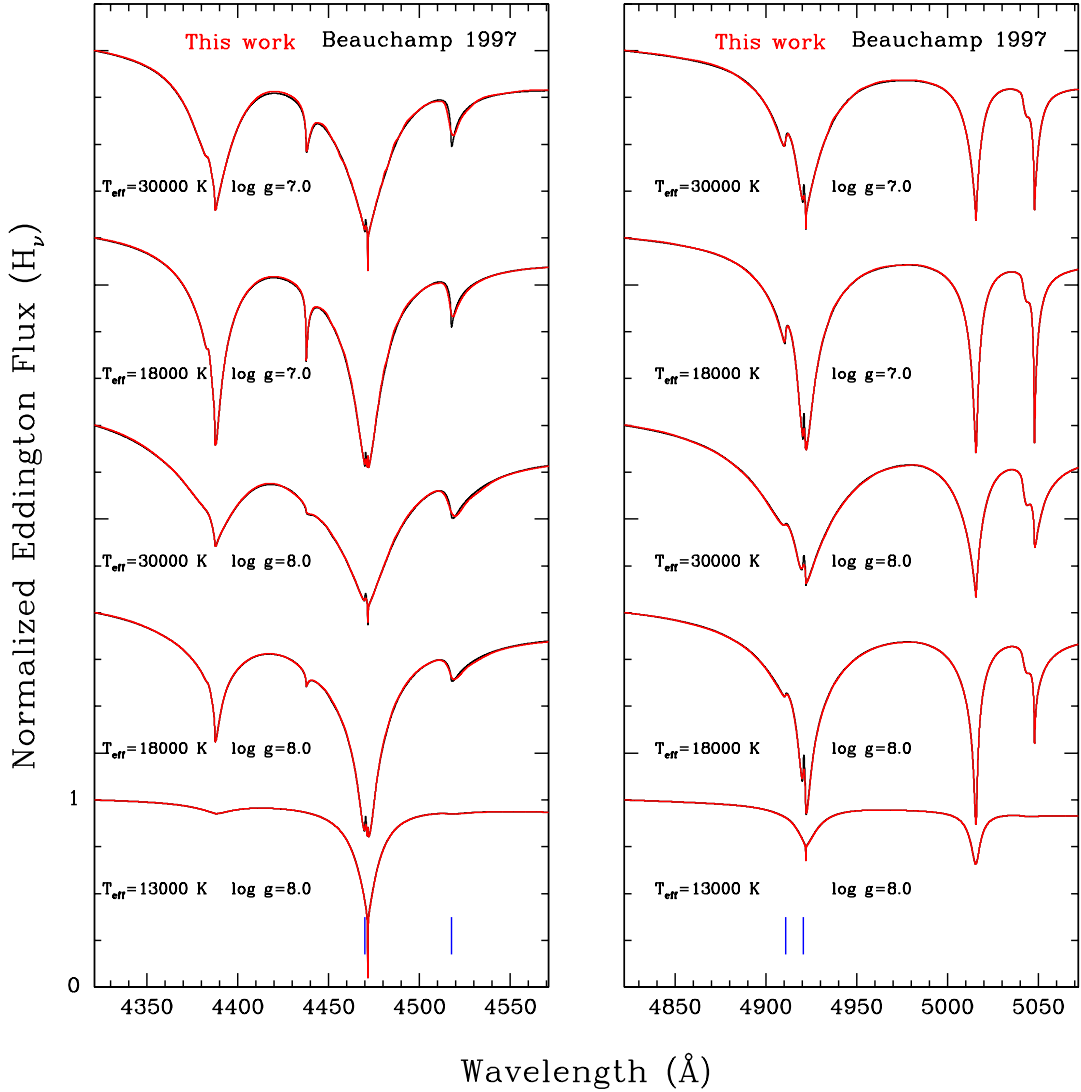}
    \caption{Comparison of synthetic spectra for DB white dwarfs near
      He {\sc i} $\lambda$4471 (left) and He {\sc i} $\lambda$4922
      (right) produced with our new Stark broadened profiles (red
      lines) and the profiles of \citet[][black
        lines]{beauchamp97}. All spectra are normalized to a continuum
      set to unity and offset from each other for clarity. The
      effective temperature and surface gravity are indicated in the
      panels; a pure helium composition is assumed throughout. The
      location of all forbidden components are indicated by blue tick
      marks at the bottom.}
    \label{fig:synthspeccomp}
\end{figure}

Since we possess a complete grid of improved profiles only for He {\sc
  i} $\lambda\lambda$4471 and 4922, we restrict our comparison of
synthetic spectra for these two lines. In Figure
\ref{fig:synthspeccomp}, we compare the (normalized) Eddington fluxes
($H_\nu$) at $\Te=13,000$~K, 18,000~K, and 30,000~K at $\logg=8.0$,
which span the range of temperatures for DB white dwarfs. Also shown
are synthetic spectra for the two hottest models at $\logg=7.0$ where
the differences are more significant. These comparisons reveal that
our improved Stark profiles yield results that are remarkably similar
to those obtained with our earlier calculations at all temperatures
where DB white dwarfs are found, at least for these two lines. The
results indicate that our computer simulations are able to produce
Stark profiles that are perfectly suited in the context of DB white
dwarfs, and also that the more approximate calculations of
\citet{beauchamp97} seem appropriate in the physical regime explored
here. This overall agreement is not unexpected given that the
temperature and electron density regime corresponds to the range of
parameters where our new Stark profiles and those of
\citet{beauchamp97} agree the best (see Section \ref{sec:comp}).

There are still subtle differences, however, between both sets of
calculations near the farthest forbidden component of He {\sc i}
$\lambda$4471, as well as near the line core; these differences are
also more significant at lower surface gravities. This is a direct
consequence of the inclusion of ion dynamics into our simulations (see
also Figure \ref{fig:prof6678} as well as
\citealt{BCS74,BCS75,Gig09,Lara12}), while these ions are assumed to
be quasi-static in the earlier calculations of
\citet{beauchamp97}. For the region near He {\sc i} $\lambda$4922, the
forbidden components are closer to the line core but similar
differences can be observed. Obviously, these effects will require
extremely high resolution and high signal-to-noise spectroscopic
observations to be measured.

Given that we have complete grids of computer simulation profiles for
only two lines, it is too early to assess the implications of our
results on the temperature and mass scales of DB white dwarfs. In
particular, we need to produce detailed Stark profiles for the He {\sc
  i} $\lambda$3820 and $\lambda$4388 lines, which are particularly
$\logg$-sensitive in the context of DB white dwarfs. Such results will
be reported in due time. Also, our computer simulations can find
applications in other astrophysical contexts as well, such as the
UV-bright star Barnard 29, recently analyzed by \citet{Dixon2019}, who
showed that ion dynamics --- using the profiles of \citet{BCS75} ---
provided a better match to the He {\sc i} $\lambda$4922 line (see
their Figure 8) over Stark profiles where these effects were
neglected.

\section{Conclusion}\label{sec:conc}

We presented new calculations of Stark broadened profiles for the He
{\sc i} $\lambda\lambda$4471 and 4922 lines using computer simulations
performed by combining the method developed by \citet{Gig96} for
treating the dynamics of the environment around the emitting neutral
helium atom, and the approach of \citet{H88} for dealing with local
density variations. These line profiles offer a better representation
of the ion dynamics than the semi-analytical calculations of
\citet{beauchamp97}, and they also represent a different view of the
computer simulations discussed by \citet{Gig09} and \citet{Lara12},
who did not consider the effect of local density variations. By also
neglecting this effect in our calculations, we were able to properly
test the validity of our computer simulation environment, by
reproducing almost perfectly the He {\sc i} $\lambda\lambda$4471 and
4922 profiles published by \citet{Gig09} and \citet{Lara12},
respectively.

An exploration of narrower helium lines, such as He {\sc i}
$\lambda\lambda$5877 and 6678, showed that the shape of the permitted
components is Lorentzian-like, while their forbidden components behave
much differently. In particular, we were able to demonstrate --- by
removing the effect of ion dynamics from our computer simulations and
by comparing the results with those of \citet{beauchamp97} --- that
the forbidden component of He {\sc i} $\lambda$6678 is significantly
affected by this effect. Also, an exploration of broader lines, such
as He {\sc i} $\lambda\lambda$4026 and 4144, showed how the CPU time
required in our simulations increases dramatically for transitions to
higher principal numbers $n$, as a result of the larger amount of
neighboring energy levels to be included in the
calculations. Fortunately, since there are no narrow features in these
profiles, some of the computing time can be reduced by decreasing the
simulation time.

Although we were able to produce profiles for a total of six neutral
helium lines so far, the computer simulation approach begins to show
its limitation regarding the amount of effort required to optimize the
quality of the line profile calculations, given the amount of memory
space and CPU time required to perform the simulation. Thus,
additional numerical challenges remain to be dealt with before we are
able to produce full grids of He {\sc i} line profiles. Nevertheless,
the results obtained so far from our computer simulations represent a
most encouraging and promising avenue of investigation.

Despite the improvements for the He {\sc i} $\lambda\lambda$4471 and
4922 line profiles obtained from our computer simulations, their
implementation into the calculations of synthetic spectra of DB white
dwarfs did not show any appreciable difference with respect to the
previous calculations of \citet{beauchamp97}. This is mostly because
the physical conditions where these lines are formed correspond to the
range where both sets of calculations agree the best. As such, our
improved calculations can be viewed as an important validation of the
approximations made by Beauchamp, at least for these two lines. A more
detailed comparison with other He {\sc i} lines will be presented in
future work.

Finally, our computer simulations should eventually provide a new
environment for exploring other physical problems encountered in the
modeling of white dwarf atmospheres, such as the Stark broadening of
hydrogen lines in DA stars, and perhaps even van der Waals broadening
in DB stars. In future work, we also expect to lift some of the
approximations discussed in this paper (see Section \ref{sec:THEO}),
namely the neglect of the mixing of lower levels, the mixing of upper
levels with levels of different $n$, and the quadrupolar term in the
multipole expansion of the potential, as carried out by
\citet{Gomez2016} for hydrogen.

\section*{Acknowledgments}
This work is supported in part by the NSERC Canada and by the Fund FRQ-NT (Québec).

\appendix
\section{Appendix: The Perturbed He {\sc i} State Basis \label{sec:per}}

Integration of the time evolution operator (Equation \ref{eq:Udyn1})
and computation of the electron-broadened profile (Equation
\ref{eq:Ie}) for a field $F$ --- whether within the impact or the
one-electron approximation --- use the perturbed basis
$|\alpha(F)\ket$, and not the set of energy states $|a\ket$ of the
isolated helium atom. Since the dipole matrix elements are only known
in the unperturbed basis, a unitary transformation must be found to
perform the change of basis. The unitary transformation and the energy
of the perturbed states are found by solving an eigenvalue
problem. Here, we consider this problem for a field ${\bf F}$ of
arbitrary orientation. The special case of a field along the $z$
direction applies to the quasi-static field from ions in the standard
theory.

The perturbed states are a solution to the following Schr\"odinger equation

\begin{equation}
\label{eq:Ha2}
    H_a({\bf F})|\alpha\rangle = E_\alpha |\alpha\rangle\ ,
\end{equation}

\noindent where $|\alpha\rangle$ corresponds to the normalized quantum
state, with energy $E_\alpha$, of the perturbed helium. The perturbed
states are projected into the subspace of $|a\ket$ unperturbed upper
states
\begin{equation}
\label{eq:alpha1}
     |\alpha\rangle = \sum_{a}\langle a|\alpha\rangle\,|a\rangle\ ,
\end{equation}

\noindent where $\langle a|\,\alpha\rangle$ are the transformation
coefficients from the unperturbed to the perturbed basis, which is
normalized such that $\langle\alpha|\alpha\rangle=1$.

By combining Equations (\ref{eq:Ha2}) and (\ref{eq:alpha1}), and by
multiplying the resulting equation by the unperturbed state set
$\langle a'|\,$, we obtain the following eigenvalue problem

\begin{equation}
\label{eq:stark6}
    \sum_{a'}(\Omega_{aa'}-E_\alpha\delta_{aa'})\,\langle a'|\,\alpha\rangle=0\ ,
\end{equation}

\noindent where 
\begin{equation}
\label{eq:Omega}
\Omega_{aa'} = 
\mathbf{F}\cdot\langle a|\mathbf{d}|a' \rangle+ E_{a}\delta_{aa'}
\end{equation}

\noindent is Hermitian, with real eigenvalues $E_\alpha$. The
eigenvectors $\bra a|\alpha\ket$ are then normalized so that $\bra
\alpha | \alpha \ket = 1$. Most elements of $\Omega$ vanish, a
consequence of the properties of the Clebsch-Gordan coefficients,
which also lead to selection rules for electric dipole transitions.

\bibliography{ms}{}

\begin{thebibliography}{77}
\expandafter\ifx\csname natexlab\endcsname\relax\def\natexlab#1{#1}\fi

\bibitem[{{Alexiou} {et~al.}(1999{\natexlab{a}}){Alexiou}, {Glenzer}, \&
  {Lee}}]{Alex99v2}
{Alexiou}, S., {Glenzer}, S., \& {Lee}, R.~W. 1999{\natexlab{a}}, \pre, 60,
  6238

\bibitem[{{Alexiou} {et~al.}(1999{\natexlab{b}}){Alexiou}, {Sauvan},
  {Poqu{\'e}russe}, {Leboucher-Dalimier}, \& {Lee}}]{Alex99}
{Alexiou}, S., {Sauvan}, P., {Poqu{\'e}russe}, A., {Leboucher-Dalimier}, E., \&
  {Lee}, R.~W. 1999{\natexlab{b}}, \pre, 59, 3499

\bibitem[{{Baranger}(1958{\natexlab{a}})}]{Baranger58c}
{Baranger}, M. 1958{\natexlab{a}}, Physical Review, 112, 855

\bibitem[{{Baranger}(1958{\natexlab{b}})}]{Baranger58a}
---. 1958{\natexlab{b}}, Physical Review, 111, 481

\bibitem[{{Baranger}(1962)}]{Baranger62}
---. 1962, Pure and Applied Physics, 13, 493

\bibitem[{{Barnard} \& {Cooper}(1970)}]{BC70}
{Barnard}, A.~J. \& {Cooper}, J. 1970, \jqsrt, 10, 695

\bibitem[{{Barnard} {et~al.}(1969){Barnard}, {Cooper}, \& {Shamey}}]{BCS69}
{Barnard}, A.~J., {Cooper}, J., \& {Shamey}, L.~J. 1969, \aap, 1, 28

\bibitem[{{Barnard} {et~al.}(1974){Barnard}, {Cooper}, \& {Smith}}]{BCS74}
{Barnard}, A.~J., {Cooper}, J., \& {Smith}, E.~W. 1974, \jqsrt, 14, 1025

\bibitem[{{Barnard} {et~al.}(1975){Barnard}, {Cooper}, \& {Smith}}]{BCS75}
---. 1975, \jqsrt, 15, 429

\bibitem[{{Beauchamp}(1995)}]{AlainPhD}
{Beauchamp}, A. 1995, PhD thesis, Université de Montréal

\bibitem[{{Beauchamp} {et~al.}(1997){Beauchamp}, {Wesemael}, \&
  {Bergeron}}]{beauchamp97}
{Beauchamp}, A., {Wesemael}, F., \& {Bergeron}, P. 1997, \apjs, 108, 559

\bibitem[{{Beauchamp} {et~al.}(1996){Beauchamp}, {Wesemael}, {Bergeron},
  {Liebert}, \& {Saffer}}]{Beauchamp1996}
{Beauchamp}, A., {Wesemael}, F., {Bergeron}, P., {Liebert}, J., \& {Saffer},
  R.~A. Astronomical Society of the Pacific Conference Series, Vol.~96, The DB
  and DBA white dwarfs: epitomes of hydrogen-deficient stars, ed. C.~S.
  {Jeffery} \& U.~{Heber}, 295

\bibitem[{{Ben Chaouacha} {et~al.}(2007){Ben Chaouacha}, {Sahal-Br{\'e}chot},
  \& {Ben Nessib}}]{SahalBrechot07}
{Ben Chaouacha}, H., {Sahal-Br{\'e}chot}, S., \& {Ben Nessib}, N. 2007, \aap,
  465, 651

\bibitem[{{Bergeron} {et~al.}(2019){Bergeron}, {Dufour}, {Fontaine}, {Coutu},
  {Blouin}, {Genest-Beaulieu}, {B{\'e}dard}, \& {Rolland }}]{Bergeron2019}
{Bergeron}, P., {Dufour}, P., {Fontaine}, G., {Coutu}, S., {Blouin}, S.,
  {Genest-Beaulieu}, C., {B{\'e}dard}, A., \& {Rolland }, B. 2019, \apj, 876,
  67

\bibitem[{{Bergeron} {et~al.}(2001){Bergeron}, {Leggett}, \& {Ruiz}}]{BLR01}
{Bergeron}, P., {Leggett}, S.~K., \& {Ruiz}, M.~T. 2001, \apjs, 133, 413

\bibitem[{{Bergeron} {et~al.}(1997){Bergeron}, {Ruiz}, \& {Leggett}}]{BRL97}
{Bergeron}, P., {Ruiz}, M.~T., \& {Leggett}, S.~K. 1997, \apjs, 108, 339

\bibitem[{{Bergeron} {et~al.}(1992){Bergeron}, {Saffer}, \&
  {Liebert}}]{Bergeron92}
{Bergeron}, P., {Saffer}, R.~A., \& {Liebert}, J. 1992, \apj, 394, 228

\bibitem[{{Bergeron} {et~al.}(2011){Bergeron}, {Wesemael}, {Dufour},
  {Beauchamp}, {Hunter}, {Saffer}, {Gianninas}, {Ruiz}, {Limoges}, {Dufour},
  {Fontaine}, \& {Liebert}}]{Bergeron2011}
{Bergeron}, P., {Wesemael}, F., {Dufour}, P., {Beauchamp}, A., {Hunter}, C.,
  {Saffer}, R.~A., {Gianninas}, A., {Ruiz}, M.~T., {Limoges}, M.~M., {Dufour},
  P., {Fontaine}, G., \& {Liebert}, J. 2011, \apj, 737, 28

\bibitem[{{Calisti} {et~al.}(1988){Calisti}, {Stamm}, \& {Talin}}]{Calisti88}
{Calisti}, A., {Stamm}, R., \& {Talin}, B. 1988, \pra, 38, 4883

\bibitem[{{Dimitrijevic} \& {Sahal-Brechot}(1984)}]{Dimitri84}
{Dimitrijevic}, M.~S. \& {Sahal-Brechot}, S. 1984, \jqsrt, 31, 301

\bibitem[{{Dixon} {et~al.}(2019){Dixon}, {Chayer}, {Reid}, \& {Miller
  Bertolami}}]{Dixon2019}
{Dixon}, W.~V., {Chayer}, P., {Reid}, I.~N., \& {Miller Bertolami}, M.~M. 2019,
  \aj, 157, 147

\bibitem[{{Eisenstein} {et~al.}(2006){Eisenstein}, {Liebert}, {Koester},
  {Kleinmann}, {Nitta}, {Smith}, {Barentine}, {Brewington}, {Brinkmann},
  {Harvanek}, {Krzesi{\'n}ski}, {Neilsen}, {Long}, {Schneider}, \&
  {Snedden}}]{Eisenstein2006}
{Eisenstein}, D.~J., {Liebert}, J., {Koester}, D., {Kleinmann}, S.~J., {Nitta},
  A., {Smith}, P.~S., {Barentine}, J.~C., {Brewington}, H.~J., {Brinkmann}, J.,
  {Harvanek}, M., {Krzesi{\'n}ski}, J., {Neilsen}, Eric~H., J., {Long}, D.,
  {Schneider}, D.~P., \& {Snedden}, S.~A. 2006, \aj, 132, 676

\bibitem[{{Ferri} {et~al.}(2014){Ferri}, {Calisti}, {Moss{\'e}}, {Rosato},
  {Talin}, {Alexiou}, {Gigosos}, {Gonz{\'a}lez}, {Gonz{\'a}lez-Herrero},
  {Lara}, {Gomez}, {Iglesias}, {Lorenzen}, {Mancini}, \&
  {Stambulchik}}]{Ferri14}
{Ferri}, S., {Calisti}, A., {Moss{\'e}}, C., {Rosato}, J., {Talin}, B.,
  {Alexiou}, S., {Gigosos}, M., {Gonz{\'a}lez}, M., {Gonz{\'a}lez-Herrero}, D.,
  {Lara}, N., {Gomez}, T., {Iglesias}, C., {Lorenzen}, S., {Mancini}, R., \&
  {Stambulchik}, E. 2014, Atoms, 2, 299

\bibitem[{{Frerichs}(1989)}]{Frerich89}
{Frerichs}, M.~R. 1989, Zeitschrift fur Physik D Atoms Molecules Clusters, 11,
  315

\bibitem[{{Gaia Collaboration} {et~al.}(2018){Gaia Collaboration}, {Babusiaux},
  {van Leeuwen}, {Barstow}, {Jordi}, {Vallenari}, {Bossini}, {Bressan},
  {Cantat-Gaudin}, {van Leeuwen}, \& et~al.}]{GaiaHRD}
{Gaia Collaboration}, {Babusiaux}, C., {van Leeuwen}, F., {Barstow}, M.~A.,
  {Jordi}, C., {Vallenari}, A., {Bossini}, D., {Bressan}, A., {Cantat-Gaudin},
  T., {van Leeuwen}, M., \& et~al. 2018, \aap, 616, A10

\bibitem[{{Genest-Beaulieu} \& {Bergeron}(2019{\natexlab{a}})}]{GenestSDSS2019}
{Genest-Beaulieu}, C. \& {Bergeron}, P. 2019{\natexlab{a}}, \apj, 871, 169

\bibitem[{{Genest-Beaulieu} \& {Bergeron}(2019{\natexlab{b}})}]{GenestDB2019}
---. 2019{\natexlab{b}}, \apj, 882, 106

\bibitem[{{Gieske} \& {Griem}(1969)}]{Gieske69}
{Gieske}, H.~A. \& {Griem}, H.~R. 1969, \apj, 157, 963

\bibitem[{{Gigosos}(2014)}]{Gig14}
{Gigosos}, M.~A. 2014, Journal of Physics D Applied Physics, 47, 343001

\bibitem[{{Gigosos} \& {Carde{\~n}oso}(1996)}]{Gig96}
{Gigosos}, M.~A. \& {Carde{\~n}oso}, V. 1996, Journal of Physics B Atomic
  Molecular Physics, 29, 4795

\bibitem[{{Gigosos} \& {Cardenoso}(1987)}]{Gig87}
{Gigosos}, M.~A. \& {Cardenoso}, V. 1987, Journal of Physics B Atomic Molecular
  Physics, 20, 6005

\bibitem[{{Gigosos} {et~al.}(1985){Gigosos}, {Fraile}, \& {Torres}}]{Gig85}
{Gigosos}, M.~A., {Fraile}, J., \& {Torres}, F. 1985, \pra, 31, 3509

\bibitem[{{Gigosos} \& {Gonz{\'a}lez}(2009)}]{Gig09}
{Gigosos}, M.~A. \& {Gonz{\'a}lez}, M.~{\'A}. 2009, \aap, 503, 293

\bibitem[{{Gigosos} {et~al.}(2003){Gigosos}, {Gonz{\'a}lez}, \&
  {Carde{\~n}oso}}]{Gig03}
{Gigosos}, M.~A., {Gonz{\'a}lez}, M.~{\'A}., \& {Carde{\~n}oso}, V. 2003,
  Spectrochimica Acta, 58, 1489

\bibitem[{{Gigosos} {et~al.}(2018){Gigosos}, {Gonz{\'a}lez-Herrero}, {Lara},
  {Florido}, {Calisti}, {Ferri}, \& {Talin}}]{Gig18}
{Gigosos}, M.~A., {Gonz{\'a}lez-Herrero}, D., {Lara}, N., {Florido}, R.,
  {Calisti}, A., {Ferri}, S., \& {Talin}, B. 2018, \pre, 98, 033307

\bibitem[{{Gomez}(2017)}]{GomezPhD}
{Gomez}, T.~A. 2017, PhD thesis, University of Texas

\bibitem[{{Gomez} {et~al.}(2016){Gomez}, {Nagayama}, {Kilcrease}, {Montgomery},
  \& {Winget}}]{Gomez2016}
{Gomez}, T.~A., {Nagayama}, T., {Kilcrease}, D.~P., {Montgomery}, M.~H., \&
  {Winget}, D.~E. 2016, \pra, 94, 022501

\bibitem[{{Griem}(1968)}]{Griem68}
{Griem}, H.~R. 1968, \apj, 154, 1111

\bibitem[{Griem(1974)}]{Griem74}
Griem, H.~R. 1974, Spectral line broadening by plasmas (New York: Academic
  Press)

\bibitem[{{Griem} {et~al.}(1962){Griem}, {Baranger}, {Kolb}, \&
  {Oertel}}]{Griem62}
{Griem}, H.~R., {Baranger}, M., {Kolb}, A.~C., \& {Oertel}, G. 1962, Physical
  Review, 125, 177

\bibitem[{{Halenka} \& {Olchawa}(1996)}]{Halenka96}
{Halenka}, J. \& {Olchawa}, W. 1996, \jqsrt, 56, 17

\bibitem[{{Halenka} {et~al.}(2002){Halenka}, {Olchawa}, {Grabowski}, \&
  {Gajda}}]{Halenka02}
{Halenka}, J., {Olchawa}, W., {Grabowski}, B., \& {Gajda}, F. 2002, \jqsrt, 74,
  539

\bibitem[{{Hegerfeldt} \& {Kesting}(1988)}]{H88}
{Hegerfeldt}, G.~C. \& {Kesting}, V. 1988, \pra, 37, 1488

\bibitem[{{Hooper}(1968)}]{Hoop68}
{Hooper}, C.~F. 1968, Physical Review, 169, 193

\bibitem[{{Hummer} \& {Mihalas}(1988)}]{Mihalas88}
{Hummer}, D.~G. \& {Mihalas}, D. 1988, \apj, 331, 794

\bibitem[{{Kepler} {et~al.}(2007){Kepler}, {Kleinman}, {Nitta}, {Koester},
  {Castanheira}, {Giovannini}, {Costa}, \& {Althaus}}]{Kepler2007}
{Kepler}, S.~O., {Kleinman}, S.~J., {Nitta}, A., {Koester}, D., {Castanheira},
  B.~G., {Giovannini}, O., {Costa}, A.~F.~M., \& {Althaus}, L. 2007, \mnras,
  375, 1315

\bibitem[{{Kepler} {et~al.}(2019){Kepler}, {Pelisoli}, {Koester}, {Reindl},
  {Geier}, {Romero}, {Ourique}, {Oliveira}, \& {Amaral}}]{Kepler2019}
{Kepler}, S.~O., {Pelisoli}, I., {Koester}, D., {Reindl}, N., {Geier}, S.,
  {Romero}, A.~D., {Ourique}, G., {Oliveira}, C. d.~P., \& {Amaral}, L.~A.
  2019, \mnras, 486, 2169

\bibitem[{{Koester} \& {Kepler}(2015)}]{Koester2015}
{Koester}, D. \& {Kepler}, S.~O. 2015, \aap, 583, A86

\bibitem[{{Kolb} \& {Griem}(1958)}]{Kolb58}
{Kolb}, A.~C. \& {Griem}, H. 1958, Physical Review, 111, 514

\bibitem[{{Lara} {et~al.}(2012){Lara}, {Gonz{\'a}lez}, \& {Gigosos}}]{Lara12}
{Lara}, N., {Gonz{\'a}lez}, M.~{\'A}., \& {Gigosos}, M.~A. 2012, \aap, 542, A75

\bibitem[{{Lorentz}(1906)}]{Lorentz1906}
{Lorentz}, H.~A. 1906, Koninklijke Nederlandse Akademie van Wetenschappen
  Proceedings Series B Physical Sciences, 8, 591

\bibitem[{{Oertel} \& {Shomo}(1968)}]{Shomo68}
{Oertel}, G.~K. \& {Shomo}, L.~P. 1968, \apjs, 16, 175

\bibitem[{{Olchawa}(2002)}]{Olcha02}
{Olchawa}, W. 2002, \jqsrt, 74, 417

\bibitem[{{Olchawa} {et~al.}(2004){Olchawa}, {Olchawa}, \&
  {Grabowski}}]{Olcha04}
{Olchawa}, W., {Olchawa}, R., \& {Grabowski}, B. 2004, European Physical
  Journal D, 28, 119

\bibitem[{{Omar} {et~al.}(2006){Omar}, {G{\"u}nter}, {Wierling}, \&
  {R{\"o}pke}}]{Omar06}
{Omar}, B., {G{\"u}nter}, S., {Wierling}, A., \& {R{\"o}pke}, G. 2006, \pre,
  73, 056405

\bibitem[{{Poqu{\'e}russe} {et~al.}(1996){Poqu{\'e}russe}, {Alexiou}, \&
  {Klodzh}}]{Poquerusse96}
{Poqu{\'e}russe}, A., {Alexiou}, S., \& {Klodzh}, E. 1996, \jqsrt, 56, 153

\bibitem[{Press {et~al.}(2007)Press, Teukolsky, Vetterling, \&
  Flannery}]{Numrecipe}
Press, W., Teukolsky, S., Vetterling, W., \& Flannery, B. 2007, Numerical
  Recipes: the art of scientific computing (Cambridge: Cambridge University
  Press)

\bibitem[{{Sahal-Brechot}(1969{\natexlab{a}})}]{Sahal69a}
{Sahal-Brechot}, S. 1969{\natexlab{a}}, \aap, 1, 91

\bibitem[{{Sahal-Brechot}(1969{\natexlab{b}})}]{Sahal69b}
---. 1969{\natexlab{b}}, \aap, 2, 322

\bibitem[{{Schoning}(1994)}]{Schoning94}
{Schoning}, T. 1994, Journal of Physics B Atomic Molecular Physics, 27, 4501

\bibitem[{{Seaton}(1990)}]{Seaton90}
{Seaton}, M.~J. 1990, Journal of Physics B Atomic Molecular Physics, 23, 3255

\bibitem[{Seidel \& Stamm(1982)}]{Seidel82}
Seidel, J. \& Stamm, R. 1982, Journal of Quantitative Spectroscopy and
  Radiative Transfer, 27, 499

\bibitem[{{Shamey}(1969)}]{Shamey69}
{Shamey}, L.~J. 1969, PhD thesis, University of Colorado at Boulder.

\bibitem[{Smith {et~al.}(1969)Smith, Vidal, \& Cooper}]{Smith69}
Smith, E., Vidal, C., \& Cooper, J. 1969, Journal of Research of the National
  Bureau of Standards, 73A

\bibitem[{{Smith} {et~al.}(1969){Smith}, {Cooper}, \& {Vidal}}]{Smith69B}
{Smith}, E.~W., {Cooper}, J., \& {Vidal}, C.~R. 1969, Physical Review, 185, 140

\bibitem[{{Sorge} \& {G{\"u}nter}(2000)}]{Sorge00}
{Sorge}, S. \& {G{\"u}nter}, S. 2000, European Physical Journal D, 12, 369

\bibitem[{{Stambulchik} {et~al.}(2007){Stambulchik}, {Fisher}, {Maron},
  {Griem}, \& {Alexiou}}]{Stambulchik2007}
{Stambulchik}, E., {Fisher}, D.~V., {Maron}, Y., {Griem}, H.~R., \& {Alexiou},
  S. 2007, High Energy Density Physics, 3, 272

\bibitem[{{Stambulchik} \& {Maron}(2006)}]{Stambulchik2006}
{Stambulchik}, E. \& {Maron}, Y. 2006, \jqsrt, 99, 730

\bibitem[{{Stamm} \& {Voslamber}(1979)}]{Stamm79}
{Stamm}, R. \& {Voslamber}, D. 1979, \jqsrt, 22, 599

\bibitem[{{Tonry} {et~al.}(2012){Tonry}, {Stubbs}, {Lykke}, {Doherty},
  {Shivvers}, {Burgett}, {Chambers}, {Hodapp}, {Kaiser}, {Kudritzki},
  {Magnier}, {Morgan}, {Price}, \& {Wainscoat}}]{Tonry2012}
{Tonry}, J.~L., {Stubbs}, C.~W., {Lykke}, K.~R., {Doherty}, P., {Shivvers},
  I.~S., {Burgett}, W.~S., {Chambers}, K.~C., {Hodapp}, K.~W., {Kaiser}, N.,
  {Kudritzki}, R.-P., {Magnier}, E.~A., {Morgan}, J.~S., {Price}, P.~A., \&
  {Wainscoat}, R.~J. 2012, \apj, 750, 99

\bibitem[{{Tremblay} \& {Bergeron}(2009)}]{Tremblay2009}
{Tremblay}, P.~E. \& {Bergeron}, P. 2009, \apj, 696, 1755

\bibitem[{{Tremblay} {et~al.}(2011){Tremblay}, {Bergeron}, \&
  {Gianninas}}]{Tremblay2011SDSS}
{Tremblay}, P.~E., {Bergeron}, P., \& {Gianninas}, A. 2011, \apj, 730, 128

\bibitem[{{Vidal} {et~al.}(1970){Vidal}, {Cooper}, \& {Smith}}]{VCS70}
{Vidal}, C.~R., {Cooper}, J., \& {Smith}, E.~W. 1970, \jqsrt, 10, 1011

\bibitem[{{Voss} {et~al.}(2007){Voss}, {Koester}, {Napiwotzki}, {Christlieb},
  \& {Reimers}}]{Voss2007}
{Voss}, B., {Koester}, D., {Napiwotzki}, R., {Christlieb}, N., \& {Reimers}, D.
  2007, \aap, 470, 1079

\bibitem[{Wujec {et~al.}(2003)Wujec, Jazgara, Halenka, \& Musielok}]{Wu03}
Wujec, T., Jazgara, A., Halenka, J., \& Musielok, J. 2003, The European
  Physical Journal D - Atomic, Molecular, Optical and Plasma Physics, 23, 405

\bibitem[{{Wujec} {et~al.}(2002){Wujec}, {Olchawa}, {Halenka}, \&
  {Musielok}}]{Wu02}
{Wujec}, T., {Olchawa}, W., {Halenka}, J., \& {Musielok}, J. 2002, \pre, 66,
  066403

\bibitem[{{Ya'akobi} {et~al.}(1972){Ya'akobi}, {George}, {Bekefi}, \&
  {Hawryluk}}]{Ya72}
{Ya'akobi}, B., {George}, E.~V., {Bekefi}, G., \& {Hawryluk}, R.~J. 1972,
  Journal of Physics B Atomic Molecular Physics, 5, 1017

\end{thebibliography}
\bibliographystyle{apj}

\end{document}